\documentclass{elsarticle}

\usepackage{acronym}
\usepackage{amsmath}
\usepackage{amssymb}
\usepackage{cleveref}
\usepackage{graphicx}
\usepackage{siunitx}
\usepackage{fullpage}

\newacro{BASIS}{Bayesian annealed sequential importance sampling}
\newacro{HMC}{Hamiltonian Monte-Carlo}
\newacro{MAP}{maximum a posteriori}
\newacro{MCMC}{Markov chain Monte-Carlo}
\newacro{TMCMC}{transitional Markov chain Monte-Carlo}
\newacro{ODIL}{optimizing a discrete loss}
\newacro{ODE}{ordinary differential equation}
\newacro{PDE}{partial differential equation}
\newacro{PINN}{physics-informed neural network}
\newacro{RBC}{red blood cell}

\newacro{MRI}{magnetic resonance imaging}
\newacro{GTV}{gross tumor volume}
\newacro{CTV}{clinical target volume}
\newacro{PTV}{planning target volume}

\DeclareMathOperator*{\argmin}{arg\,min}
\DeclareMathOperator*{\argmax}{arg\,max}

\graphicspath{{figures/}{.}}
\newcommand{\figbf}[1]{\uppercase{\textbf{#1}}}

\title{Bayesian Inference for PDE-based Inverse Problems \\using the Optimization of a Discrete Loss}
\date{}

\author[1]{Lucas Amoudruz}
\author[1]{Sergey Litvinov}
\author[2]{Costas Papadimitriou}
\author[1]{Petros Koumoutsakos}
\affiliation[1]{Computational Science and Engineering Laboratory, Harvard John A. Paulson School of Engineering and Applied Sciences, 19 Oxford St, Cambridge, MA 02138, United States.}
\affiliation[2]{System Dynamics Laboratory, Department of Mechanical Engineering, University of Thessaly, Volos, 38334, Greece}

\begin{document}

\begin{abstract}
  Inverse problems are crucial for many applications in science, engineering and medicine that involve data assimilation, design, and  imaging.
  Their solution infers the parameters or latent states of a complex system from noisy data and partially observable processes.
  When measurements are an incomplete or indirect view of the system, additional knowledge is required to accurately solve the inverse problem.
  Adopting a physical model of the system in the form of partial differential equations (PDEs) is a potent method to close this gap.
  In particular, the method of optimizing a discrete loss (ODIL) has shown great potential in terms of robustness and computational cost.
  In this work, we introduce B-ODIL, a Bayesian extension of ODIL, that integrates the PDE loss of ODIL as prior knowledge and combines it with a likelihood describing the data.
  B-ODIL employs a Bayesian formulation of PDE-based inverse problems to infer solutions with quantified uncertainties.
  We demonstrate the capabilities of B-ODIL in a series of synthetic benchmarks involving PDEs in one, two, and three dimensions.
  We showcase the application of B-ODIL in estimating tumor concentration and its uncertainty in a patient's brain from MRI scans using a three-dimensional tumor growth model.
\end{abstract}


\maketitle

\section{Introduction}

Inverse problems are ubiquitous in science, engineering, and medicine, in particular for problems where observations provide only indirect or incomplete information about a system~\citep{Ghattas2021Learning}.
Inverse problems are central in a wide range of applications such as flow field reconstruction~\citep{gunes2008use,ding2023full,karnakov2023flow}, data assimilation~\citep{carrassi2018data}, medical imaging~\citep{suetens2017fundamentals,jin2017deep}, and parameters estimation of material properties~\citep{angelikopoulos2012bayesian,economides2021hierarchical,amoudruz2023stress}.
A particularly challenging class of inverse problems arises when the forward model is governed by \acp{ODE} or \acp{PDE}~\citep{isakov2006inverse}.
Incorporating physical knowledge through mathematical models constrains the space of admissible solutions and can reduce the severity of ill-posedness, although inverse problems often remain ill-posed and still require regularization~\citep{Galbally2010Nonlinear,Lieberman2010Parameter,stuart2010inverse}.
However, this approach can suffer from the high dimensionality of the problem, stiffness, noisy measurements, and sensitivity to parameters.
In particular, quantifying the uncertainties of solutions is challenging with standard techniques for inverse \ac{PDE} problems such as Bayesian inference~\citep{tarantola2005inverse,stuart2010inverse}, variational methods~\citep{scherzer2009variational}, ensemble Kalman methods~\citep{iglesias2016regularizing}, and adjoint-based optimization~\citep{van2015penalty}, which can be limited with issues of scalability, robustness, and computational cost.
Bayesian inference for PDE-constrained inverse problems, Laplace and Gauss-Newton-type approximations, has been extensively studied in the context of function-space formulations~\citep{stuart2010inverse,bui2013computational,petra2014computational,antil2024efficient}.

In parallel, operator learning approaches based on DeepONets~\citep{lu2021learning}, Fourier neural operators~\citep{li2020fourier}, and graph neural networks~\citep{gao2022physics,duthe2025graph} have been extended to inverse problems and uncertainty quantification~\citep{zou2024neuraluq,molinaro2023neural,kaltenbach2022semi}.
Similar Bayesian techniques rely on training data to build prior knowledge~\citep{adler2018deep}.
However, the application of these operator learning techniques to large-scale problems is limited by the cost of their training and the difficulty of generating sufficient high-fidelity data.
In addition, their performance can degrade when the training data do not cover the regimes encountered in the inverse problem, making their generalization challenging.

More recently, \ac{PDE}-based inverse problems were solved with the methods of \acp{PINN}~\citep{raissi2019physics} and \ac{ODIL}~\citep{karnakov2024solving}.
In both these approaches, the solution of the inverse problem is obtained by minimizing a loss that contains two terms: the deviation between the field and the data, and the residuals of the \ac{PDE} evaluated at collocation points in space-time.
Combining these terms into a single loss was pioneered by Leeuwen and Herrmann and applied to linear \acp{PDE}~\citep{van2015penalty}.
\Acp{PINN} and \ac{ODIL} differ in their representation of the field and in the way \ac{PDE} residuals are estimated.
\Acp{PINN} represent the field as the output of a neural network that has space-time as its input.
Residuals of the \ac{PDE} are then estimated through automatic differentiation.
In contrast, fields in \ac{ODIL} are stored on a grid, and \ac{PDE} residuals are estimated using traditional discretizations, leading to a considerable computational advantage over \acp{PINN} because of the locality of these operators~\citep{karnakov2024solving}.
\Acp{PINN} and \ac{ODIL} have been successful in many applications ranging from fluid mechanics~\citep{cai2021physics,karnakov2023flow,buhendwa2025data,scott2024using} to tumor growth~\citep{balcerak2023individualizing,balcerak2024physics} and learning policies for fluid control and manipulation~\citep{karnakov2025optimal,amoudruz2025contactless}.

Despite these advances, \acp{PINN} and \ac{ODIL} solutions can be affected by measurement errors of the provided data.
In particular, it is not clear how these measurement errors affect the uncertainties of the solution of the inverse problem.
Recent studies have addressed these issues in the context of \acp{PINN}~\citep{yang2021b,nair2025pinns} through a Bayesian and a conformal prediction framework~\citep{yu2025conformal} approach to account for the variability of the unknown field.
However, a similar theory has not been developed for \ac{ODIL}.
We note that the Bayesian extensions of \acp{PINN} have been applied to problems of less than two dimensions.
On the other hand, a Bayesian extension of \ac{ODIL} could potentially provide quantified uncertainties for inverse problems in higher dimensions, as \ac{ODIL} was shown to be orders of magnitude faster and more robust than \acp{PINN} in two and three dimensional benchmark problems~\citep{karnakov2024solving}.

In this study, we present B-ODIL, a Bayesian extension of \ac{ODIL}.
B-ODIL enables uncertainty quantification and robustness to model error in high-dimensional PDE-based inverse problems, while retaining the computational structure and scalability of ODIL.
In this framework, the prior incorporates knowledge from the \ac{PDE}, while the likelihood couples the observed data to the unknown field.
This method provides a solution to the inverse problem with quantified uncertainties.
We account for the computational burden by estimating the posterior distribution with different sampling techniques and two approximation strategies: (i) a full Laplace approximation to the joint posterior for moderate-dimensional problems, and (ii) a parameter-focused mode approximation designed for scalable inference in large-scale cases.
We have designed a series of benchmarks with increasing levels of complexity to test B-ODIL.
First, we consider the \ac{ODE} describing the dynamics of a harmonic oscillator, to compare the validity of the Laplace and mode approximations compared with \ac{HMC} sampling of the same posterior in a computationally tractable setting.
We then apply B-ODIL to the one dimensional \ac{PDE} of the diffusion equation with unknown initial conditions, introducing the challenge of ill-posed inverse problems, typical in PDE-based inference, where uncertainty quantification is crucial.
The third benchmark tests the method's ability to reconstruct the states of a non-linear two dimensional \ac{PDE} from synthetic data, and we demonstrate that the ground truth falls within the uncertainty bounds provided by B-ODIL.
Finally, we apply the method to a three dimensional model of tumor growth coupled with real patients data, and provide estimates of tumor cell fields with quantified uncertainties given medical images.

\section{Methods}
\label{se:methods}

\subsection{ODIL}
\label{se:ODIL}

We consider a \ac{PDE} defined in the space-time domain $\Omega$ with boundaries and initial conditions on $\partial \Omega$,
\begin{equation} \label{eq:pde}
  \begin{aligned}
    \mathcal{L}(\mathbf{u}, \mathbf{\theta}) &= 0, & \mathrm{in} \; \Omega, \\
    \mathcal{B}(\mathbf{u}, \mathbf{\theta}) &= 0, & \mathrm{on} \; \partial \Omega,
  \end{aligned}
\end{equation}
where $\mathcal{L}$ is a differential operator that encodes the \ac{PDE}, $\mathcal{B}$ encodes the boundary and initial conditions, $\mathbf{u}$ is the unknown field and $\mathbf{\theta}$ are the parameters of the model.
Discretizing over space and time into a grid, we can solve the discrete version of \cref{eq:pde}:
\begin{equation} \label{eq:pde:discrete}
  \begin{aligned}
    \mathcal{L}^h_i(\mathbf{u}, \mathbf{\theta}) &= 0, & i=1,\dots,N, \\
    \mathcal{B}^h_j(\mathbf{u}, \mathbf{\theta}) &= 0, & j=1,\dots,N_B,
  \end{aligned}
\end{equation}
where $\mathcal{L}^h$ and $\mathcal{B}^h$ are the discretized versions of $\mathcal{L}$ and $\mathcal{B}$, respectively, and $\mathbf{u}$ is the unknown discretized field.
To solve the discretized problem, we can reformulate \cref{eq:pde:discrete} as a minimization of the loss
\begin{equation} \label{eq:odil:nodata:loss}
  L_\mathrm{PDE}(\mathbf{u}, \mathbf{\theta}) = \frac{1}{N} \sum\limits_{i=1}^{N} \mathcal{L}^h_i(\mathbf{u}, \mathbf{\theta})^2 + \frac{1}{N_B} \sum\limits_{j=1}^{N_B} \mathcal{B}^h_j(\mathbf{u}, \mathbf{\theta})^2,
\end{equation}
where $N$ and $N_B$ are the number of discrete components of the discretized operators $\mathcal{L}^h$ and $\mathcal{B}^h$, respectively.
In the setting of an inverse problem, parts of the problem description such as initial conditions and boundary conditions may be missing and need to be estimated from measurements.
Given these measurements $\{y_k\}_{k=1}^{N_D}$ and the corresponding measurement operators $h_k(\mathbf{u}, \mathbf{\theta})$, $k=1,2\dots N_D$, the inverse problem can be solved by minimizing the loss
\begin{equation} \label{eq:odil:loss}
  L(\mathbf{u}, \mathbf{\theta}) = L_\mathrm{PDE}(\mathbf{u}, \mathbf{\theta}) + \frac{\lambda}{N_D} \sum\limits_{k=1}^{N_D} \left( y_k - h_k(\mathbf{u}, \mathbf{\theta})\right)^2,
\end{equation}
where $\lambda$ is a positive constant that describes the importance of fitting the data with respect to satisfying the discrete \ac{PDE}.
In this work, the minimization of \cref{eq:odil:loss} is performed with gradient-based methods implemented in Pytorch~\citep{paszke2019pytorch} and detailed in the results section.

\subsection{B-ODIL: Bayesian inference for inverse problems with ODIL}
\label{se:BODIL}

We now consider the Bayesian formulation of \ac{PDE}-based inverse problems, where the goal is to infer parameters $\theta$ and the solution field $\mathbf{u}$ from noisy measurements $\mathcal{D} = \{y_k\}_{k=1}^{N_D}$, given a model in the form of a \ac{PDE}.
In this framework, the posterior distribution is obtained by combining a likelihood model for the data with prior knowledge enforcing compatibility between the solution field and the \ac{PDE}.
As we have seen in the previous section, \ac{ODIL} provides a natural way to encode this compatibility through a \ac{PDE} loss.
We thus incorporate this \ac{PDE} loss into the prior information, allowing us to extend \ac{ODIL} into the Bayesian setting.
In classical Bayesian inverse problems with well-posed PDEs and accurate forward models, one often eliminates the state variable via a map $\mathbf{u}=f(\mathbf{\theta})$, yielding a reduced posterior over $\mathbf{\theta}$.
In contrast, B-ODIL targets settings in which the governing equations are enforced approximately through a residual-based term and may be affected by discretization error or model misspecification.
In this regime, enforcing a strict parameter-to-state map can lead to biased and overconfident posteriors, motivating the use of a joint posterior over $(\mathbf{u}, \mathbf{\theta})$ that can accommodate structural model error.
We assume that the measurements $\mathcal{D}$ are noisy and that we have a model describing this noise.
We would like to estimate the parameters $\mathbf{\theta}$ and the solution $\mathbf{u}$, with quantified uncertainties.
According to the Bayes' theorem, the posterior distribution of these quantities given the data reads
\begin{equation}
  P(\mathbf{u}, \mathbf{\theta} \mid \mathcal{D}) \propto P(\mathcal{D} \mid \mathbf{u}, \mathbf{\theta}) P(\mathbf{u}, \mathbf{\theta}),
\end{equation}
where we omit the normalization constant $P(\mathcal{D})$.
The term $P(\mathbf{u}, \mathbf{\theta})$ denotes the prior knowledge of the parameters and solution field, and $P(\mathcal{D} \mid \mathbf{u}, \mathbf{\theta})$ is the likelihood of the data.
A common form of the likelihood assumes that the observations are statistically independent and that they are normally distributed:
\begin{equation} \label{eq:likelihood}
  P(\mathcal{D}\mid\mathbf{u}, \mathbf{\theta}) = \prod\limits_{k=1}^{N_D} \frac{1}{\sqrt{2\pi\sigma^2}} \exp\left( - \frac{(y_k - h_k(\mathbf{u}, \mathbf{\theta}))^2}{2\sigma^2} \right),
\end{equation}
where $\sigma$ is the standard deviation associated with the observable and $h_k$ , $k=1,\dots, N_D$, are measurement operators on the solution $\mathbf{u}$.
For example, $h_k$ can be a linear interpolation of the field at the measurement location from the discrete field $\mathbf{u}$.
Throughout this work, measurements are assumed conditionally independent given $(\mathbf{u},\mathbf{\theta})$, corresponding to independent observation noise; correlations between measurements could be incorporated through a non-diagonal likelihood covariance but are not considered here for simplicity.
Other likelihood models can be used within the same framework.
Examples are provided in the results section.

As prior knowledge, we assume that the possible solutions $\mathbf{u}$ are compatible with the \ac{PDE} associated with the model.
To account for this knowledge, the prior over the solution $\mathbf{u}$ with parameters $\mathbf{\theta}$ is such that the loss $L_\mathrm{PDE}(\mathbf{u},\mathbf{\theta})$ is small, and thus we choose priors of the form
\begin{equation} \label{eq:full:prior}
  P(\mathbf{u}, \mathbf{\theta}) = \frac{1}{Z} \exp\left( -\beta L_\mathrm{PDE}(\mathbf{u},\mathbf{\theta}) \right) P(\mathbf{\theta}),
\end{equation}
where $\beta$ is a positive scalar that controls how peaked the distribution is, and $Z$ is a normalization constant that does not depend on $\mathbf{u}$ or $\mathbf{\theta}$.
Finally, $P(\mathbf{\theta})$ contains additional prior knowledge on $\mathbf{\theta}$.
In this work, $P(\mathbf{\theta})$ is taken as improper uniform prior in all cases.
The prior in \cref{eq:full:prior} is defined at the discrete level through the loss $L_\mathrm{PDE}$.
As a result, the posterior is a discretization-dependent object.
In this work we adopt a computational perspective and evaluate posterior stability under mesh refinement (see \cref{se:oscillator}), which provides an empirical check that discretization effects are negligible once posterior summaries have converged.
We refer to \citet{stuart2010inverse,cotter2013mcmc} for the function-space formulation of Bayesian inverse problems and discretization-robust algorithms.
The posterior distribution of the solution and parameters of the model thus becomes
\begin{equation}
  P(\mathbf{u}, \mathbf{\theta} | \mathcal{D}) \propto P(\mathcal{D} | \mathbf{u}, \mathbf{\theta}) \frac{1}{Z} \exp\left( -\beta L_\mathrm{PDE}(\mathbf{u},\mathbf{\theta}) \right) P(\mathbf{\theta}).
\end{equation}
The dimensionality of this problem  is large due to the unknown variables $\mathbf{u}$, challenging sampling techniques.
To obtain tractable uncertainty estimates in high dimensions, we use the Laplace approximation to the joint posterior.
The log-posterior reads
\begin{equation} \label{eq:odil:posterior}
  \log P(\mathbf{u}, \mathbf{\theta} | \mathcal{D}) = \log P(\mathcal{D} | \mathbf{u}, \mathbf{\theta})
  -\beta L_\mathrm{PDE}(\mathbf{u},\mathbf{\theta}) - \log{Z}  + \log P(\mathbf{\theta}) + C,
\end{equation}
where $C$ is a scalar that does not depend on $\mathbf{\theta}$ or $\mathbf{u}$.
We expand this quantity in Taylor series up to second order around the \ac{MAP}, where the linear term vanishes, leading to the Laplace approximation~\citep{tierney1986accurate}
\begin{equation} \label{eq:odil:posterior:laplace}
  \log P(\mathbf{v} | D) \approx \log P(\mathbf{v}^\star | D) + \frac{1}{2} (\mathbf{v} - \mathbf{v}^\star)^T H (\mathbf{v} - \mathbf{v}^\star),
\end{equation}
where $\mathbf{v} = (\mathbf{u}, \mathbf{\theta})$, $\mathbf{v}^\star$ is the solution at the \ac{MAP}, and $H$ is the Hessian of the log-posterior evaluated at $\mathbf{v}^\star$.
Thus, the posterior is approximated with a multivariate Gaussian with mean $\mathbf{v}^\star$ and covariance $\Sigma = -H^{-1}$.
As with any local Gaussian approximation, the Laplace method is expected to be inaccurate for strongly non-Gaussian or multimodal posteriors.
In such regimes, full sampling or alternative approximations would be required.
We note that when the likelihood takes the form of \cref{eq:likelihood}, maximizing \cref{eq:odil:posterior}, i.e.\ finding the \ac{MAP}, corresponds to the original \ac{ODIL} method, \cref{eq:odil:loss} with $\lambda = N_D / \beta$, assuming that $P(\mathbf{u})$ and $P(\mathbf{\theta})$ are taken as improper uniform priors.
Thus, computing uncertainties over predictions that were computed with ODIL consists only in computing the Hessian and inverting it.
In general, forming and inverting the Hessian with respect to $\mathbf{v}=(\mathbf{u},\mathbf{\theta})$ is computationally prohibitive when $\mathbf{u}$ is high-dimensional; consequently, the full Laplace approximation is feasible only for relatively low-dimensional benchmarks, while large-scale problems require parameter-focused or reduced approximations discussed below.

\subsection{Inference of model parameters for inverse problems}
\label{se:BODIL:params}

In the previous section we formulated  a posterior distribution based on the prior knowledge that contains information about the \ac{PDE} (\cref{eq:full:prior}) and on the likelihood of observing the data (\cref{eq:likelihood}) that fits well with the ODIL formulation.
This results in a joint posterior distribution for the unknown field $\mathbf{u}$ and the model parameters $\mathbf{\theta}$.
This distribution typically lies on a large-dimensional space, and it becomes quickly intractable to sample from this distribution.
Similarly, in large dimensions, obtaining the Laplace approximation is costly, as the computation of the Hessian of the posterior increases quadratically with the size of the problem.
In some cases, we are only interested in the posterior distribution of the model parameters, which can be obtained by marginalizing over $\mathbf{u}$:
\begin{equation} \label{eq:posterior:marginal:theta}
  P(\mathbf{\theta} | \mathcal{D}) = \int P(\mathbf{u}, \mathbf{\theta} | \mathcal{D})  d \mathbf{u}.
\end{equation}
The evaluation of this high-dimensional integral is intractable.
Instead, we assume that the joint distribution is peaked around the \ac{MAP} $\mathbf{u}^\star(\mathbf{\theta}) = \argmax\limits_{\mathbf{u}} P(\mathbf{u} | \mathbf{\theta}, \mathcal{D})$ and we approximate the posterior distribution of the model parameters using the mode approximation~\citep{tierney1986accurate}
\begin{equation} \label{eq:posterior:params:approx:MAP}
  P(\mathbf{\theta} | \mathcal{D}) \approx P(\mathbf{u}^\star(\mathbf{\theta}), \mathbf{\theta} | \mathcal{D}).
\end{equation}
In practice, the dimensionality of $\mathbf{\theta}$ is small compared to that of $\mathbf{u}$.
Thus, sampling from \cref{eq:posterior:params:approx:MAP} can be performed with traditional sampling methods such as \ac{MCMC} or \ac{TMCMC}, where each sample involves an optimization problem with respect to $\mathbf{u}$.

The formulation adopted here is not intended as a replacement for classical Bayesian inference based on repeated forward model evaluations under a strict parameter-to-state map $\mathbf{u}=f(\mathbf{\theta})$.
Such reduced formulations are appropriate when the forward model is accurate and can be evaluated efficiently for each value of $\mathbf{\theta}$.
In contrast, B-ODIL targets settings in which the governing equations may be misspecified.
We note that several studies have approached this problem by directly altering the model~\citep{morrison2018representing} or by statistically correcting its output~\citep{wang2023stochastic}.
Here, we instead incorporate the \ac{PDE} residual as a soft probabilistic constraint in the prior.
Embedding optimization over $\mathbf{u}$ within inference over $\mathbf{\theta}$ allows the PDE constraints and the data to be reconciled at the field level, while the parameter $\beta$ controls tolerance to model error.
As shown in \cref{se:oscillator:beta}, this is essential for obtaining calibrated posteriors under misspecification, where standard Bayesian inference can otherwise become overconfident and biased.

\subsection{Selection of the parameter $\beta$}
\label{se:methods:beta}

The parameter $\beta$ controls the relative importance of the PDE-residual prior and the data likelihood.
When the governing equations are known to be accurate, large values of $\beta$ enforce strong adherence to the model.
The limit $\beta\rightarrow\infty$ corresponds to classical methods for PDE-based Bayesian inverse problems.
In the presence of model error or misspecification, however, overly large $\beta$ can lead to biased and overconfident posterior estimates.

In practice, $\beta$ can be selected in different ways depending on the availability of data and computational resources.
A principled approach consists in selecting $\beta$ by maximizing the predictive log-likelihood on held-out data.
Specifically, the available observations are split into training and validation sets.
For each candidate value of $\beta$, B-ODIL is applied to the training data to obtain a posterior distribution, which is then evaluated on the validation data.
The value $\beta^\star$ maximizing the validation log-likelihood is retained.

When validation is impractical (e.g., limited data or high computational cost), we select $\beta$ by tuning, guided by the same trade-off as in classical ODIL and by posterior diagnostics (e.g., predictive uncertainty and residual structure).
Alternatively, $\beta$ can be treated as a hyperparameter and selected by maximizing the model evidence, at the cost of introducing a prior on $\beta$ and additional approximation or optimization steps.
In the present work we primarily use cross-validation when feasible, and otherwise adopt pragmatic tuning that is consistent with ODIL practice and supported by posterior predictive checks.

\section{Results}
\label{se:results}

We demonstrate the applicability of this method  by first illustrating its application to simple problems such as the Harmonic Oscillator and the diffusion equation.
We then proceed to examine the formulation for the  reaction-diffusion equations and then its  application to estimate uncertainties of a tumor concentration field given observed \ac{MRI} data of real patients.

\subsection{Harmonic oscillator}
\label{se:oscillator}

\begin{figure}
  \centering
  \includegraphics{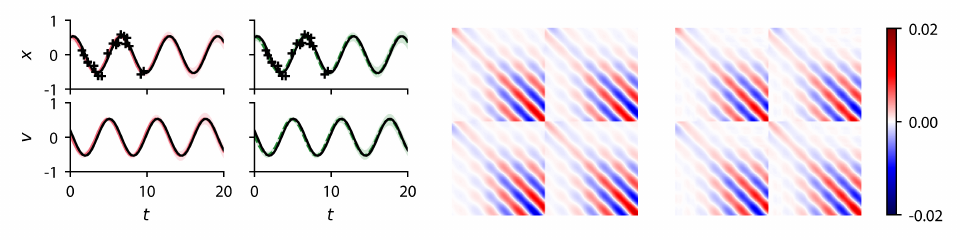}
  \caption{Left: Prediction of the position (top row) and velocity (bottom row) of the oscillator given the data (crosses) using the B-ODIL framework with Laplace approximation (left column) and HMC (right column).
    The shaded area denotes the 5 to 95\% quantiles of the posterior, and the solid line denotes the posterior mean.
    The dashed line represents the underlying process that was used to generate the data.
    Right: Covariance matrices of the full solution $(x_1,\dots, x_N, v_1, \dots, v_N)$ obtained from the Laplace approximation (left column) and estimated from HMC samples (right column).}
  \label{fig:oscillator}
\end{figure}

We first consider a simple example to demonstrate the validity of the Laplace approximation and that of \cref{eq:posterior:params:approx:MAP}.
Consider the system of \acp{ODE} of a harmonic oscillator with position $x$, velocity $v$, mass $m$ and spring coefficient $k$,
\begin{align}
  \frac{dx}{dt} &= v, \\
  \frac{dv}{dt} &= -\frac{k}{m} x,
\end{align}
for time $t \in (0, T)$ and with unknown initial conditions.
Given measurements of position at known times $\mathcal{D} = \left\{t_j, x_j\right\}_{j=1}^{N_D}$ with known standard deviation $\sigma$, we want to predict the position and velocity of the system throughout the whole time interval.
We discretize the position and velocity into $N=64$ equidistant time intervals with $\Delta t = T / N$ and $T=20$.
The discrete losses are computed using the midpoint rule,
\begin{equation} \label{eq:oscillator:linear}
  \mathcal{L}^h_{i}(\mathbf{x}, \mathbf{v}) = \left(\frac{x_{i+1} - x_i}{\Delta t} - \frac{v_{i+1} + v_i}{2},  \frac{v_{i+1} - v_i}{\Delta t} + \frac{k}{m} \frac{x_{i+1} + x_i}{2}\right).
\end{equation}
Finally, the observation functions correspond to linear interpolations of the positions at data points $t_j$, $j=1,2,\dots,N_D$.
For now, we assume that the parameter $\omega = \sqrt{k/m} = 1$ is known.
We use synthetic data consisting of $N_D = 20$ time points uniformly distributed in $t\in(0, T/2)$, and corresponding positions with Gaussian distribution centered around the exact solution of the model with standard deviation of $\sigma=0.1$.
The initial conditions used to generate the data were set to $x(0) = 0.5$ and $v(0) = 0.2$.
We choose a value of $\beta = \SI{e4}{}$ in the prior distribution (see \cref{eq:full:prior}).
The \ac{MAP} is computed by maximizing the posterior with respect to $(\mathbf{x}, \mathbf{v})$ with the Adam optimizer~\citep{kingma2014adam} with learning rate $\SI{e-3}{}$ for 10'000 epochs.
Computing the Laplace approximation for this case takes about $\SI{2}{\second}$ on a consumer-grade laptop.
The Laplace approximation of this problem is shown on \cref{fig:oscillator}.
For comparison, we solve the same problem with \ac{HMC} on the same posterior, and show the results on the same figure.
The \ac{HMC} results were obtained from a custom implementation of the original \ac{HMC} algorithm~\citep{neal2011mcmc} with 10'000 samples, chain lengths $l=10$, mass matrix $M=1$, and step size $\delta t = 0.008$, tuned for an acceptance rate of $\alpha \approx 0.65$.
This implementation with these parameters took about $\SI{5}{\minute}$ on a consumer-grade laptop.

With both Laplace and \ac{HMC} approaches, the uncertainties increase away from the data that was used to calibrate the model.
Furthermore, these uncertainties are similar between both methods, despite the Gaussian approximation in the Laplace approach.
To compare the methods, we show the correlation matrix of the entries in the discretized solution $(\mathbf{x}, \mathbf{v})$ on \cref{fig:oscillator}.
In the Laplace approach, this quantity corresponds to the inverse of the Hessian matrix of the log-posterior distribution.
With \ac{HMC}, we estimate this matrix by computing the empirical correlation between samples.
The correlation matrices obtained from the two methods are close to each other, indicating that the Laplace approximation gives a good representation of the posterior distribution.

\begin{figure}
  \centering
  \includegraphics{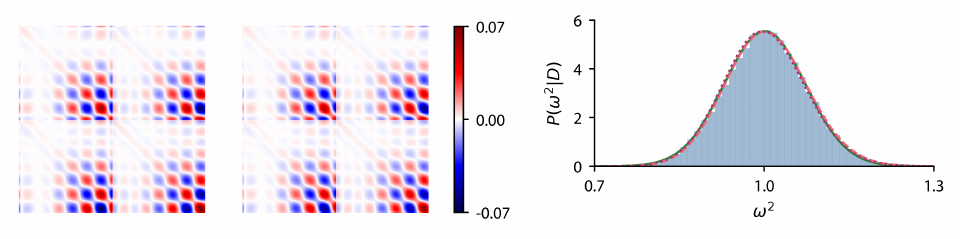}
  \caption{Left: Covariance matrices of the posterior distribution of the full solution $(\omega^2, x_1,\dots, x_N, v_1, \dots, v_N)$ obtained with the Laplace and HMC methods.
    Right:
    Marginal posterior probability of $\omega^2$ obtained from HMC (histogram), Laplace (solid line), mode approximation given by \cref{eq:posterior:params:approx:MAP} (dashed line), and exact solution (dots).
  }
  \label{fig:oscillator:omega}
\end{figure}

We also consider the same problem but with an unknown value of $\omega$.
The procedure for the Laplace approach and \ac{HMC} sampling are the same as above except that now we also infer the model parameter $\mathbf{\theta} = \omega^2 = k/m$.
The covariance matrices of the vector $(\omega^2, \mathbf{x}, \mathbf{v})$ are shown in \cref{fig:oscillator:omega} and have a very similar structure.
In addition, we estimate the marginal distribution of the model parameter $\omega$ using the approximation given by \cref{eq:posterior:params:approx:MAP}, evaluated at 50 equally spaced points for $0.7 \leq \omega \leq 1.3$.
The marginal distribution of $\omega^2$ is shown in \cref{fig:oscillator:omega}.
For the Laplace approximation, this distribution corresponds to a Gaussian with mean around the \ac{MAP} $(\omega^\star)^2$ and variance $(H^{-1})_{\omega^2,\omega^2}$.
The \ac{HMC}, Laplace and mode approximations show excellent agreement with the exact solution, centered around the reference value $\omega^2 = 1$ that was used to generate the data.
The exact solution is obtained by expanding \cref{eq:posterior:marginal:theta}, noting that the logarithm of the joint density of $\mathbf{u}, \mathbf{\theta}$ is quadratic in $\mathbf{u}$ for a fixed $\mathbf{\theta}$.
Thus, for a fixed $\mathbf{\theta} = \omega^2$,
\[
\log P(\mathbf{u}, \mathbf{\theta} | \mathcal{D}) =  l(\mathbf{u}^\star(\mathbf{\theta}), \mathbf{\theta}) + \frac 1 2 (\mathbf{u} - \mathbf{u}^\star(\mathbf{\theta}))^T H(\mathbf{\theta}) (\mathbf{u} - \mathbf{u}^\star(\mathbf{\theta})),
\]
where $l$ is the log-likelihood, $\mathbf{u}^\star(\mathbf{\theta}) = \argmax\limits_{\mathbf{u}} l(\mathbf{u}, \mathbf{\theta})$, and $H(\mathbf{\theta})$ is the Hessian matrix of $l$ with respect to $\mathbf{u}$, evaluated at $\mathbf{u}^\star(\mathbf{\theta})$.
Replacing this expression into \cref{eq:posterior:marginal:theta}, we get
\[
P(\mathbf{\theta} | \mathcal{D}) =  \frac{(2\pi)^{d/2}}{\sqrt{\det  H(\mathbf{\theta})} } \exp l(\mathbf{u}^\star(\mathbf{\theta}), \mathbf{\theta}),
\]
where $d$ is the dimension of $\mathbf{u}$.
To evaluate this expression for each value of $\theta$, we use the optimal value $\mathbf{u}^\star(\mathbf{\theta})$ estimated with \ac{ODIL}, and use automatic differentiation in Pytorch to compute $H(\mathbf{\theta})$.
We use this approach throughout this work.
In practice, $\mathbf{u}^\star(\mathbf{\theta})$ is obtained by numerical optimization and is therefore only approximate; the resulting expression should be interpreted as a Laplace-type approximation based on a locally quadratic expansion around a numerically converged solution.

\begin{figure}
  \centering
  \includegraphics{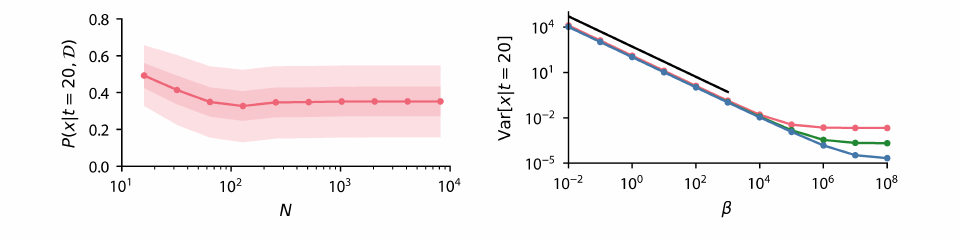}
  \caption{
    Left: Prediction of the position $x$ at time $t=20$ with the Laplace approximation against the grid size $n_t$.
    Symbols represent the mean prediction and shaded areas represent the 50\% and 90\% confidence intervals.
    Right: Variance of the prediction of $x$ at time $t=20$ against the parameter $\beta$, with $N_D=10, 100$, and $1000$, from top to bottom.
  The solid black line shows a linear decrease with $\beta$.}
  \label{fig:oscillator:conv:beta}
\end{figure}

Finally, we study the effect of the grid size $N$ and parameter $\beta$.
First we fix the \ac{PDE} weight $\beta = \SI{e4}{}$ and the number of measurements $N_D = 20$, and predict the position $x$ at the last grid point ($t=20$) given the data and the discretized \ac{PDE}.
\Cref{fig:oscillator:conv:beta} shows that the posterior of $x(t=20)$ converges as we increase the grid size.
In practice, and throughout this work, we select the smallest grid resolution for which posterior summaries have converged, in order to minimize computational cost while avoiding discretization-induced artifacts.

Second, we vary the parameter $\beta$ for three values of $N_D$ and with a fixed resolution $N=256$, for which the posterior has converged.
The variance of the posterior prediction of $x$ at time $t=20$ is shown against $\beta$ in \cref{fig:oscillator:conv:beta}.
For each value of $N_D$, we observe two regimes.
For low values of $\beta$, corresponding to soft priors on the model, the variance decreases linearly with $\beta$.
When $\beta$ is large, the variance reaches a plateau that is limited by the uncertainties coming from the data term.
As expected, a larger amount of data $N_D$ decreases the value of this plateau.
In the limit $\beta\rightarrow \infty$, the discretized \ac{PDE} is enforced exactly, and thus the only uncertainties come from the data.
Empirically, larger $\beta$ values increase the conditioning of the optimization problem and the number of iterations required for convergence, without improving predictions once the posterior variance has reached the data-limited plateau.
We thus use values of $\beta$ that are smaller than the start of the plateau.
From a Bayesian perspective, $\beta$ controls the tolerance to model error by tempering the influence of the PDE-based prior relative to the data likelihood.
In many practical cases, we do not know models that are perfectly consistent with observations, in which case a lower value of $\beta$ is favored.
We give more details about this choice in the next section.

\subsection{Choice of the parameter $\beta$ under model misspecification}
\label{se:oscillator:beta}

We now illustrate the practical impact of the $\beta$-selection strategy introduced in \cref{se:methods:beta} in a setting with deliberate model misspecification.
Specifically, we consider a variation of the harmonic oscillator example from \cref{se:oscillator}, in which the data are generated by an unknown nonlinear oscillator while inference is performed using the linear model.
The nonlinear oscillator follows the dynamics
\begin{align}
  \frac{dx}{dt} &= v, \\
  \frac{dv}{dt} &= -\frac{1}{m} (k_1 x + k_2 x^3),
\end{align}
where $k_1$ and $k_2$ are positive constants.
We generate $N_D=200$ data points using $x_0 = 2$, $v_0 =0$, $m=15$, $k_1=1$, $k_2=2$, and noise $\sigma=0.4$.
We then apply the B-ODIL framework with the linear model described by \cref{eq:oscillator:linear}, using the Adam optimizer with a learning rate of $\SI{5e-4}{}$ over 50000 epochs.
This procedure mimics many practical cases where the statistical model is an approximation of the true underlying data generation process.
We split the data into training and validation sets (80\% and 20\% of the data, respectively), run B-ODIL on the training set with different values of $\beta$, and select $\beta^\star$ by maximizing the validation log-likelihood.

\begin{figure}
  \centering
  \includegraphics{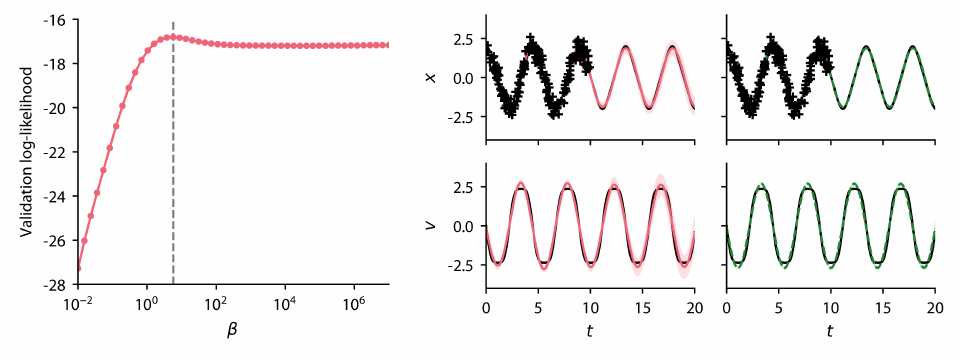}
  \caption{
    Predictions of and oscillator dynamics with a linear spring model with data coming from a nonlinear spring.
    Left: log-likelihood of the validation set against the parameter $\beta$.
    We select the value $\beta^\star$ that maximizes this log-likelihood.
    Right: Posterior prediction of the position and velocity field with the B-ODIL framework with $\beta=\beta^\star$ (left column), and with conventional Bayesian inference (right column).
    The solid black line shows the exact underlying process used to produce the data.}
  \label{fig:oscillator:nl:beta:comp}
\end{figure}

\Cref{fig:oscillator:nl:beta:comp} shows the log-likelihood of the validation set against $\beta$.
High values of $\beta$ result in a plateauing log-likelihood, as the method tends to conventional Bayesian inference as $\beta \rightarrow \infty$.
On the other hand, low values of $\beta$ lead to a decrease of the log-likelihood as the model is less and less trusted to make sharp predictions.
This log-likelihood has a maximum around $\beta^\star \approx 5.7$, which balances fitting the data and trusting the model.
We use this value to make predictions and show the posterior distribution of the position and the velocity in \cref{fig:oscillator:nl:beta:comp}.
The 90\% confidence intervals of the posterior contain the true underlying process.
In contrast, conventional Bayesian inference based on the misspecified linear model produces overly concentrated posteriors that fail to account for structural model error, leading to confidence intervals that do not capture the true dynamics away from the data (\cref{fig:oscillator:nl:beta:comp}).
This result was obtained with the MCMC sampling method with 100'000 samples and a proposal step size tuned to have a 35\% acceptance rate.

\subsection{Diffusion equation}
\label{se:diffusion}

\begin{figure}
  \centering
  \includegraphics{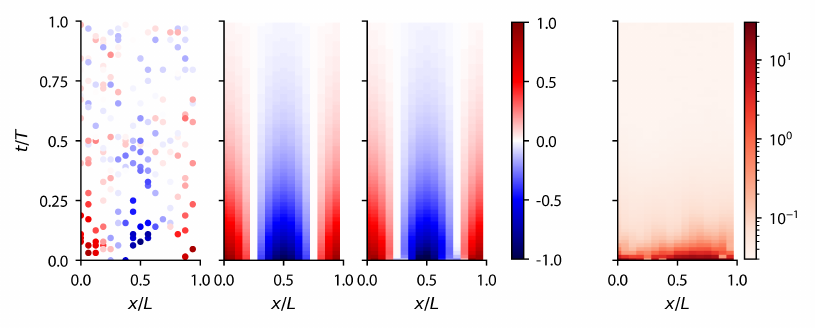}
  \caption{Laplace approximation applied to the diffusion equation.
    From left to right: Data used for inferring the field; Exact solution, used to generate the data; MAP solution of the diffusion problem given the data; spread of uncertainty given by the Laplace approximation (5-95\% quantiles).}
  \label{fig:diffusion}
\end{figure}

We now consider the one-dimensional diffusion equation with a known diffusion coefficient $D=0.1$, described by the \ac{PDE}
\begin{equation} \label{eq:diffusion:1D}
  \frac{\partial u}{\partial t} - D \frac{\partial^2 u}{\partial x^2} = 0, \quad \text{on}\quad (0, L) \times (0, T),
\end{equation}
with periodic boundary conditions and unknown initial conditions.
We generate a synthetic dataset of $N_D = 200$ measurements of the field $u$ at uniformly sampled locations in space-time, $\mathcal{D} = \{x_i. t_i, u_i\}_{i=1}^{N_D}$, with Gaussian noise of known magnitude $\sigma=0.1$ (\cref{fig:diffusion}).
The data is generated with initial conditions $u(x, 0) = \cos (2\pi x / L)$.
We want to infer the field $u$ given this dataset $\mathcal{D}$ and \cref{eq:diffusion:1D}.

We first discretize \cref{eq:diffusion:1D} with finite differences on a uniform grid,
\begin{equation} \label{eq:odil:diffusion}
  \mathcal{L}^h(\mathbf{u})_i^n = \frac{u_{i}^{n+1} - u_{i}^{n}}{\Delta t} - D \frac{u_{i-1}^{n+\frac{1}{2}} - 2 u_{i}^{n+\frac{1}{2}} + u_{i+1}^{n+\frac{1}{2}}}{\Delta x^2} = 0,
\end{equation}
where $u_i^n \approx u(x_i, t_n)$ with $x_i = i \Delta x$ and $t_n = n \Delta t$.
Furthermore, we have defined $u_{i}^{n+\frac{1}{2}} = \left(u_i^{n+1} + u_i^{n} \right) / 2$, $\Delta x = L / n_x$, and $\Delta t = T / n_t$.
We set $L=1$, $n_x = 16$, $T=1$ and $n_t = 64$.
Following \cref{eq:odil:posterior}, the posterior distribution of $\mathbf{u}$ is given by
\begin{equation} \label{eq:odil:diffusion:posterior}
  \log P(\mathbf{u} | \mathcal{D}) =
  -\sum\limits_{k=1}^{N_D} \frac{(u_k - \mathbf{u}_{i_k}^{n_k})^2}{2\sigma^2} -\frac{N_D}{2} \log {2\pi\sigma^2}
  -\frac{\beta}{N} \sum\limits_{i=1}^{N} (\mathcal{L}^h_i(\mathbf{u}))^2,
\end{equation}
where we have set $\beta=\SI{e4}{}$.
We compute the \ac{MAP} by maximizing \cref{eq:odil:diffusion:posterior} using the \texttt{LBFGS} optimizer~\citep{liu1989limited} as implemented in PyTorch.
Each call to the optimizer performs up to 20 internal quasi-Newton iterations with a strong Wolfe line search.
We run 500 such optimization steps, retaining a curvature history of size 100 throughout.
The initial step size is set to $\SI{5e-4}{}$.
Computing the Laplace approximation for this case takes about $\SI{10}{\second}$ on a consumer-grade laptop.
This inverse problem is ill-posed because the initial conditions are unknown and the diffusion process is inherently not time reversible~\citep{isakov2006inverse}.
Thus, many different initial states can produce indistinguishable noisy measurements at later times.
The \ac{MAP} associated with \cref{eq:odil:diffusion} is shown on \cref{fig:diffusion}, and we can observe deviations between the inferred field at $t=0$ and the initial conditions used to generate the dataset.
This deviation is expected at early times, as explained above.

In order to obtain uncertainties of $\mathbf{u}$, we apply the Laplace method to \cref{eq:odil:diffusion:posterior}.
We note that since \cref{eq:odil:diffusion:posterior} is quadratic in $\mathbf{u}$, the Laplace method is exact in this case.
\Cref{fig:diffusion} shows the predicted uncertainties of the field $u$.
At time $t=0$, the uncertainties are large compared to the magnitude of the field.
This is again expected since inferring initial conditions is an ill-posed problem in this case.
In contrast, at larger times, the uncertainties over $u$ are much lower (about \SI{3e-2}{}), consistent with the fact that data at previous times reduced the range of possible values at larger times. We remark that in this 1024-dimensional space setting, sampling  with \ac{HMC} was unsuccessful.
The solution with the Laplace approach is exact so \ac{HMC} sampling was not needed, but this suggests that in the next sections, where the dimension of the problems are much larger, \ac{HMC} sampling is unlikely to be successful.
Thus, we ignore this approach in the rest of this study.

\subsection{Reaction-diffusion equation}
\label{se:reaction:diffusion}

\begin{figure}
  \centering
  \includegraphics{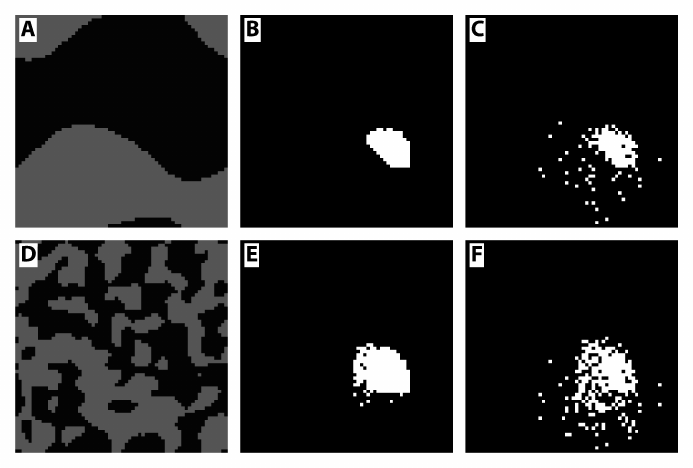}
  \caption{Reaction diffusion data.
    (\figbf{a},\figbf{d}) Diffusion field $D(x, y)$ for cases 1 and 2, respectively black and grey regions have values $D=0.005$ and $D=0.1$, respectively.
    (\figbf{b},\figbf{c}) Data with $\sigma=0.01$ and $\sigma=0.05$, respectively, for case 1.
    (\figbf{e},\figbf{f}) Data with $\sigma=0.01$ and $\sigma=0.05$, respectively, for case 2.
  }
  \label{fig:reaction:diffusion:setup}
\end{figure}

We consider the reaction-diffusion \ac{PDE} on the time-space domain $\Omega = [0, L] \times [0, L] \times [0, T]$,
\begin{equation} \label{eq:reaction:diffusion:PDE}
  \frac{\partial u}{\partial t} = \nabla \cdot (D \nabla u) + \rho u (1-u),
\end{equation}
with periodic boundary conditions in space, where $D(x, y)$ is the diffusion coefficient, $\rho$ is the reaction rate, and $u: \Omega \rightarrow [0, 1]$ the concentration field.
The diffusion coefficient is generated by thresholding a random field with filtered Gaussian frequencies in Fourier space.
We consider two cases: one with low frequency modes (case 1) and the one with higher frequencies (case 2), see \cref{fig:reaction:diffusion:setup}.

A dataset $\mathcal{D} = \{y_{ij}\}_{i,j=1}^{n_x, n_y}$ was generated from measurements of the synthetic field at the final time.
Each measurement is assumed to be statistically independent from the others and follows the binomial distribution
\begin{equation} \label{eq:reaction:diffusion:model:data}
  P(y_{ij} | u_{ij}) = \alpha_{ij}^{y_{ij}} \cdot (1 - \alpha_{ij})^{1 - y_{ij}},
\end{equation}
where
\begin{equation} \label{eq:reaction:diffusion:model:alpha}
  \alpha_{ij} = S\left(\frac{u_{ij} - \tau}{\sigma}\right), \quad S(x) = \frac{1}{1 + e^{-x}},
\end{equation}
where $\tau$ is a threshold value set to $\tau=0.5$ and $\sigma$ the scale of measurement errors.
We consider cases with $\sigma=0.01$ and $\sigma=0.1$, and show the corresponding data on \cref{fig:reaction:diffusion:setup}.

We want to infer the initial conditions $u(x, y, 0)$ given this data, and reconstruct the whole concentration field on $\Omega$ with uncertainties on the initial conditions.
We parameterize the initial conditions as
\begin{equation} \label{eq:reaction:diffusion:IC}
  u(x, y, 0) = \exp \left( -\frac{(x-x_0)^2 + (y-y_0)^2}{2 R^2} \right),
\end{equation}
where $x_0$ an $y_0$ are the position and $R$ the radius of the initial concentration field.
We set $R = L/16$ and want to infer the initial position $\mathbf{\theta} = (x_0, y_0)$ with quantified uncertainties using \cref{eq:posterior:params:approx:MAP}.

The log-likelihood of the data is given by
\begin{equation}
  \log P(\mathcal{D} | \mathbf{u}) = \sum\limits_{i=1}^{n_x}\sum\limits_{j=1}^{n_y} \log P(y_{ij} | u_{ij}).
\end{equation}
The dataset $\mathcal{D}$ is synthetically generated for each case using \cref{eq:reaction:diffusion:model:data} with the numerical solution of \cref{eq:reaction:diffusion:PDE} at time $T$, with initial conditions described by \cref{eq:reaction:diffusion:IC}.
The solution is obtained with  $\mathbf{\theta}_\mathrm{ref} = (2L/3, L/3)$, $L=1$, $T=0.5$, and $\rho=8$.
The diffusion coefficient takes values $D(x, y) \in \{0.005, 0.1\}$.
Finally, we use $n_x = n_y = 64$ points along each spatial dimension and a time step $\Delta t = \SI{6.1e-4}{}$.

\begin{figure}
  \centering
  \includegraphics{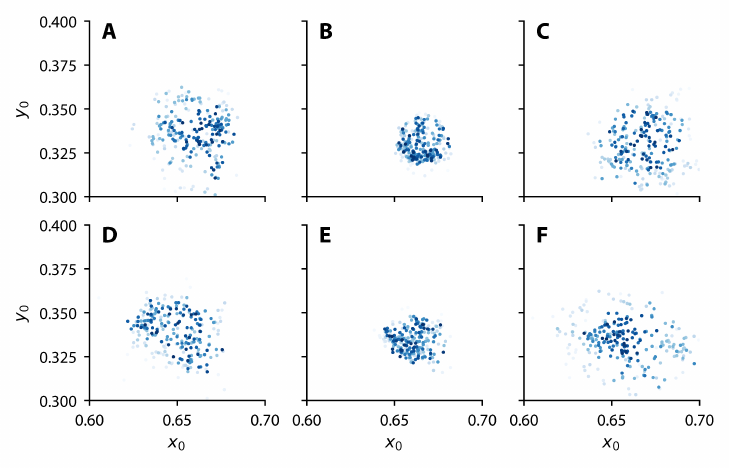}
  \caption{Samples from the posterior distribution of the initial conditions $x_0$, $y_0$ obtained with \ac{TMCMC}.
    Colors indicate the rank of the log-posterior value of each sample.
    (\figbf{a}) case 1, $\sigma=0.05$, $\lambda_\mathrm{PDE} = 10$, $\lambda_\mathrm{IC}=100$.
    (\figbf{b}) case 1, $\sigma=0.01$, $\lambda_\mathrm{PDE} = 100$, $\lambda_\mathrm{IC}=1000$.
    (\figbf{c}) case 1, $\sigma=0.01$, $\lambda_\mathrm{PDE} = 10$, $\lambda_\mathrm{IC}=100$.
    (\figbf{d}) case 2, $\sigma=0.05$, $\lambda_\mathrm{PDE} = 10$, $\lambda_\mathrm{IC}=100$.
    (\figbf{e}) case 2, $\sigma=0.01$, $\lambda_\mathrm{PDE} = 100$, $\lambda_\mathrm{IC}=1000$.
    (\figbf{f}) case 2, $\sigma=0.01$, $\lambda_\mathrm{PDE} = 10$, $\lambda_\mathrm{IC}=100$.
  }
  \label{fig:reaction:diffusion:samples}
\end{figure}

\begin{figure}
  \centering
  \includegraphics{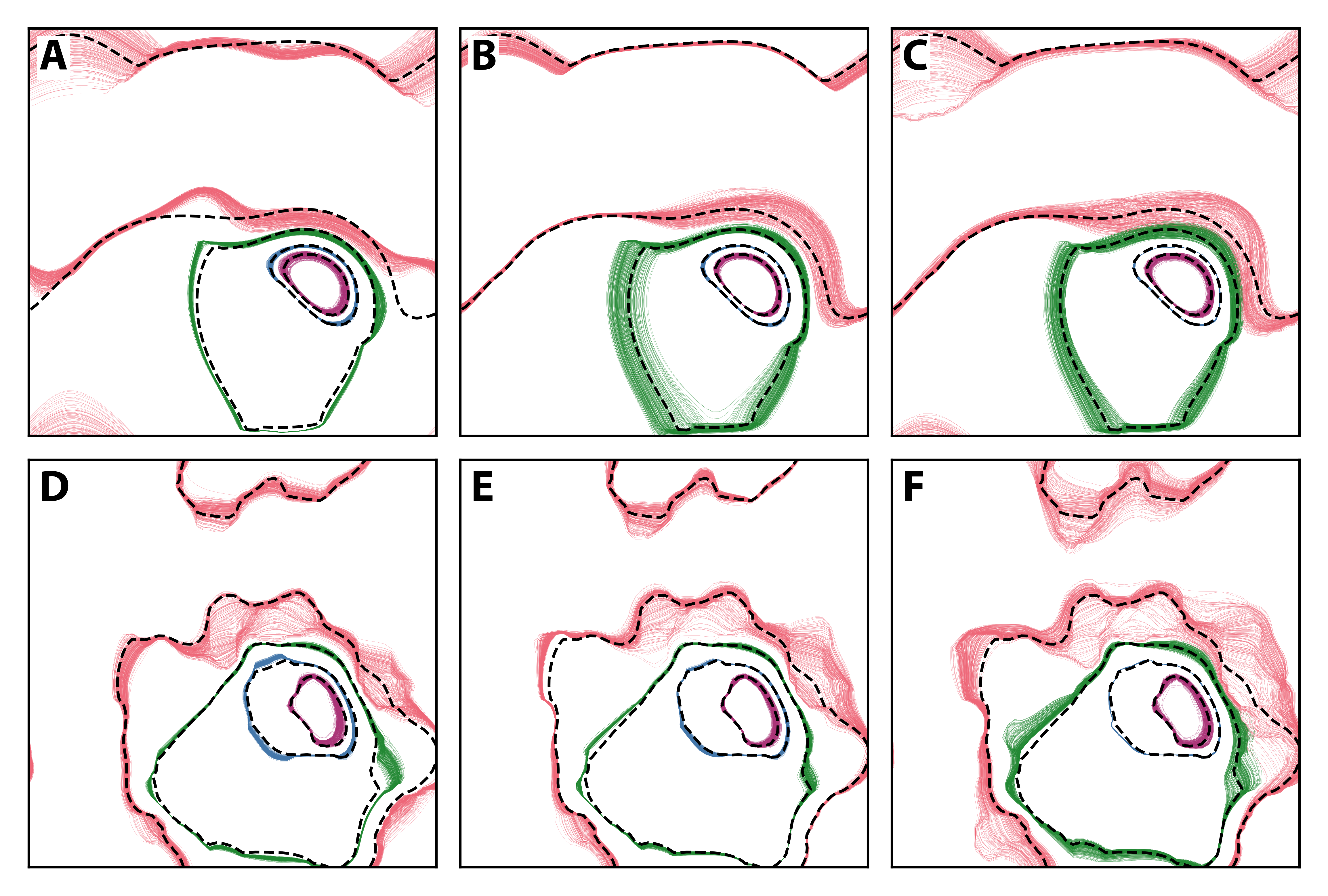}
  \caption{Reconstructed contours of the concentration field from 256 samples of the posterior distribution.
    Same cases as in \cref{fig:reaction:diffusion:samples}.
    The lines corresponds to isocontours of 0.1, 0.3, 0.5 and 0.6 for red, green, blue and magenta, respectively.
  }
  \label{fig:reaction:diffusion:results}
\end{figure}

The \ac{ODIL} solution uses the same spatial resolution as the forward solver, but we use $n_t = 129$ points along the time dimension.
The discretization of the \ac{PDE} employs the midpoint rule in time, and second order finite differences in space.
Combining the likelihood, the \ac{PDE} loss and the loss constraining the initial conditions, the log-posterior of this problem is given by
\begin{equation} \label{eq:reaction:diffusion:logposterior}
  \log P(\mathbf{u},\mathbf{\theta} | \mathcal{D}) = \lambda_\mathrm{PDE} n_x n_y n_t L_\mathrm{PDE}(\mathbf{u},\mathbf{\theta}) + \lambda_\mathrm{IC} n_x n_y L_\mathrm{IC}(\mathbf{u},\mathbf{\theta}) + \log P(\mathcal{D} | \mathbf{u}),
\end{equation}
where $L_\mathrm{IC}(\mathbf{u},\mathbf{\theta})$ is the mean squared error between $\mathbf{u}$ at time 0 and \cref{eq:reaction:diffusion:IC} evaluated at the grid points.
Finally, we use $\lambda_\mathrm{PDE} = 10$ and $\lambda_\mathrm{IC} = 100$ for $\sigma=0.05$, and $\lambda_\mathrm{PDE} = 100$ and $\lambda_\mathrm{IC} = 1000$ for $\sigma=0.01$.

We use the \ac{BASIS} algorithm~\citep{wu2018bayesian} to sample the posterior distribution of $\mathbf{\theta} = (x_0, y_0)$, with parameters $\beta_\text{BASIS}=0.2$, $\gamma_\text{BASIS} = 1$, $l_{max}=1$ and 256 samples.
Each sample is computed with 20000 epochs of the Adam optimizer with learning rate $\SI{e-3}{}$ and required approximately $\SI{8}{\minute}$ on a consumer-grade laptop and $\SI{2.3}{\minute}$ on a A100 GPU.
The posterior distribution of the initial position $\mathbf{\theta} = (x_0, y_0)$ given the data is computed for different values of $\sigma$ in both cases 1 and 2.
Samples of these distributions are shown in \cref{fig:reaction:diffusion:samples}.
A higher value of $\sigma$ means less trust in the data, hence a higher uncertainty in $\mathbf{\theta}$.
Using samples from these posterior distributions, we can reconstruct the field $u_f$ at $t=T$ and evaluate their uncertainties by using
\begin{equation}
  P(u_f|\mathcal{D}) = \int \delta\left(g(\mathbf{\theta}, \mathcal{D}) - u_f\right) P(\mathbf{\theta} | \mathcal{D}) d\mathbf{\theta} \approx \frac{1}{K}\sum\limits_{k=1}^{K} \delta\left(g(\mathbf{\theta}_k, \mathcal{D}) - u_f\right), \;\; \mathbf{\theta}_k \sim P(\mathbf{\theta} | \mathcal{D}),
\end{equation}
where $P(\mathbf{\theta} | \mathcal{D})$ is the posterior distribution of $\mathbf{\theta}$ given the data $\mathcal{D}$, and $g(\mathbf{\theta},\mathcal{D})$ is the \ac{ODIL} solution of the inverse problem.
Contours of the probability density of the concentration field at $t=T$ are shown on \cref{fig:reaction:diffusion:results}.
In both cases, the predictions are consistent with the ground-truth values, and the uncertainties in the contours are larger when the confidence in the data is lower, i.e.\ when $\sigma$ is larger.
Furthermore, the reconstruction of the 0.5 contours have a much lower uncertainty than the contours at 0.1 levels.
This is consistent with the data being measured around a threshold $\tau = 0.5$, thus the inferred concentration field at time $T$ has much more spatial information around these values.

\subsection{Reconstruction of tumor cells concentration in the brain}
\label{se:gliodil}

Medical imaging such as \ac{MRI} can detect regions on high tumor concentration in the brain, known as \ac{GTV}.
However, regions with lower tumor cell density, often containing microscopic signatures of a disease, are not detected using standard imaging modalities.
Accurately estimating the full tumor concentration field is essential for developing effective radiation therapy.
Recently, a large body of work has focused on biophysical tumor growth models and their use for patient-specific inference from medical imaging.
These approaches typically rely on reaction-diffusion or anisotropic diffusion models, sometimes coupled with mass effect, and have been used for tasks such as tumor growth prediction, parameter estimation, and image-based personalization
(see e.g., \citet{clatz2005realistic,hogea2007robust,hogea2008image,swanson2002quantifying,biegler2010large,lipkova2019personalized,menze2014multimodal,miniere2025data,lamonica2025investigating} and references therein).
The newly introduced method of GliODIL~\citep{balcerak2023individualizing} uses \ac{ODIL} to estimate the tumor concentration satisfying a reaction-diffusion \ac{PDE} while matching \ac{MRI} data of brain tumors.

Here we extend this work to estimate uncertainties of the tumor concentration field given observed \ac{MRI} data of real patients.
We then design a \ac{CTV} based on the estimated concentration field.
To estimate the concentration field at the time corresponding to the \ac{MRI} acquisition (time $T$), GliODIL estimates the density field across the whole time interval $[0, T]$, assuming a small localized tumor at time 0, and assuming that it follows the reaction-diffusion equation
\begin{equation} \label{eq:gliodil:PDE}
  \frac{\partial u}{\partial t} = \nabla \cdot (D \nabla u) + \rho u (1-u),
\end{equation}
with zero flux at the boundaries of the brain $\nabla u \cdot \mathbf{n} = 0$ on $\partial \Omega$, $\mathbf{n}$ being the normal vector at the brain boundaries.
The no-flux condition is enforced by setting the diffusion coefficient $D$ to zero outside brain tissue, so that diffusive fluxes vanish at the interface between brain matter and surrounding regions.
Furthermore we assume initial conditions of the form
\begin{equation} \label{eq:gliodil:IC}
  u(x, y, z, 0) = \begin{cases}
    0, & u_0(x, y, z) \leq 0.1, \\
    1, & u_0(x, y, z) \geq 1, \\
    u_0(x, y, z), & \text{otherwise},
  \end{cases}
\end{equation}
where
\begin{equation}
  u_0(x, y, z) = \frac{M}{(4 \pi D_t^2)^{3/2}} \exp\left( -\frac{(x-x_0)^2 + (y-y_0)^2 + (z-z_0)^2}{4 D_t^2} \right),
\end{equation}
with fixed values $M=1500$ and $D_t^2=\SI{15}{\milli\meter\squared}$ and initial tumor position $(x_0, y_0, z_0)$.
Furthermore we set the diffusion coefficient as
\begin{equation}
  D(x, y, z) = D_g c_g(x, y, z) + D_w c_w(x, y, z),
\end{equation}
where $c_g$ and $c_w$ are the proportion of gray and white matter, respectively, and $D_w$ and $D_g$ are diffusion coefficient parameters.

The data corresponds to the segmentation of \ac{MRI} scans from patients with brain tumors, where each voxel in the brain tissue is classified as belonging to the necrotic core, glioma, or healthy tissue.
The data model relates the concentration field $u$ to these classes, and we define the log-likelihood as (see also \cref{app:gliodil:loglike}):
\begin{equation} \label{eq:gliodil:loglike}
  \mathcal{L}(\mathcal{D}; \mathbf{u}, \mathbf{\theta}) = \frac{1}{\sigma} \sum\limits_{i,j,k} \min \left(u_{ijk} - \tau_{ijk, \text{lo}}, 0\right) + \min \left(0, \tau_{ijk, \text{up}} - u_{ijk}\right),
\end{equation}
where $\sigma = 0.05$ and $u_{ijk}$ is the concentration at voxel indexed by $i,j,k$.
The lower and upper thresholds $\tau_{ijk, \text{lo}}$ and $\tau_{ijk, \text{up}}$ define an interval for the concentration at each voxel, and are constructed as
\begin{equation}
  (\tau_{ijk, \text{lo}}, \tau_{ijk, \text{up}}) =
  \begin{cases}
    (0, \tau_\text{lo}), & \text{if voxel} \; $ijk$  \; \text{is classified as ``healthy''}, \\
    (\tau_\text{lo}, \tau_\text{up}), & \text{if voxel} \; $ijk$  \; \text{is classified as ``glioma''}, \\
    (\tau_\text{up}, 1), & \text{if voxel} \; $ijk$  \; \text{is classified as ``necrotic core''},
  \end{cases}
\end{equation}
where $0 < \tau_\text{lo} < \tau_\text{up} < 1$ are parameters.
The solution $u$ is discretized on a uniform grid that spans a space 50\% larger than that spanned by the tumor at time $T$ with $64\times 64 \times 64$ grid points in space and 128 points in time.
\Cref{eq:gliodil:PDE} is discretized using the Crank-Nicolson scheme, as explained in \citet{balcerak2023individualizing} and in \cref{app:gliodil}.

In addition to the tumor concentration field $u(x, y, z, t)$ over time and space, the parameters of the model and of the initial conditions are inferred: the diffusion coefficient parameters $D_w$ and $D_g$ in white and grey matter, respectively; the reaction rate $\rho$; the initial tumor position $(x_0, y_0, z_0)$; and the threshold values $\tau_\text{lo}$ and $\tau_\text{up}$.

\begin{figure}
  \centering
  \includegraphics{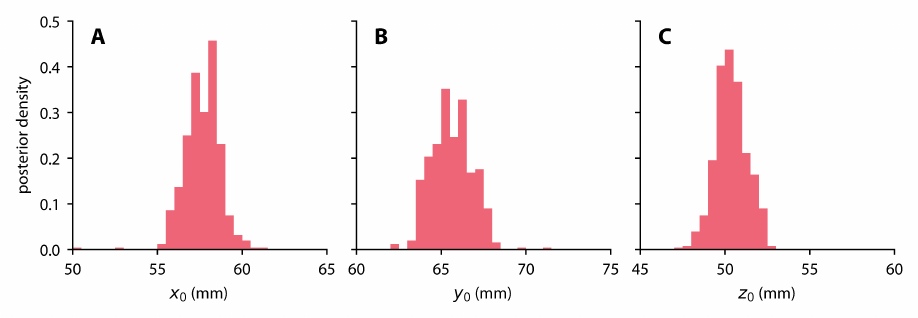}
  \caption{Marginal posterior distribution of the tumor initial position, given the data.}
  \label{fig:gliodil:posterior}
\end{figure}

\begin{figure}
  \centering
  \includegraphics[width=\textwidth]{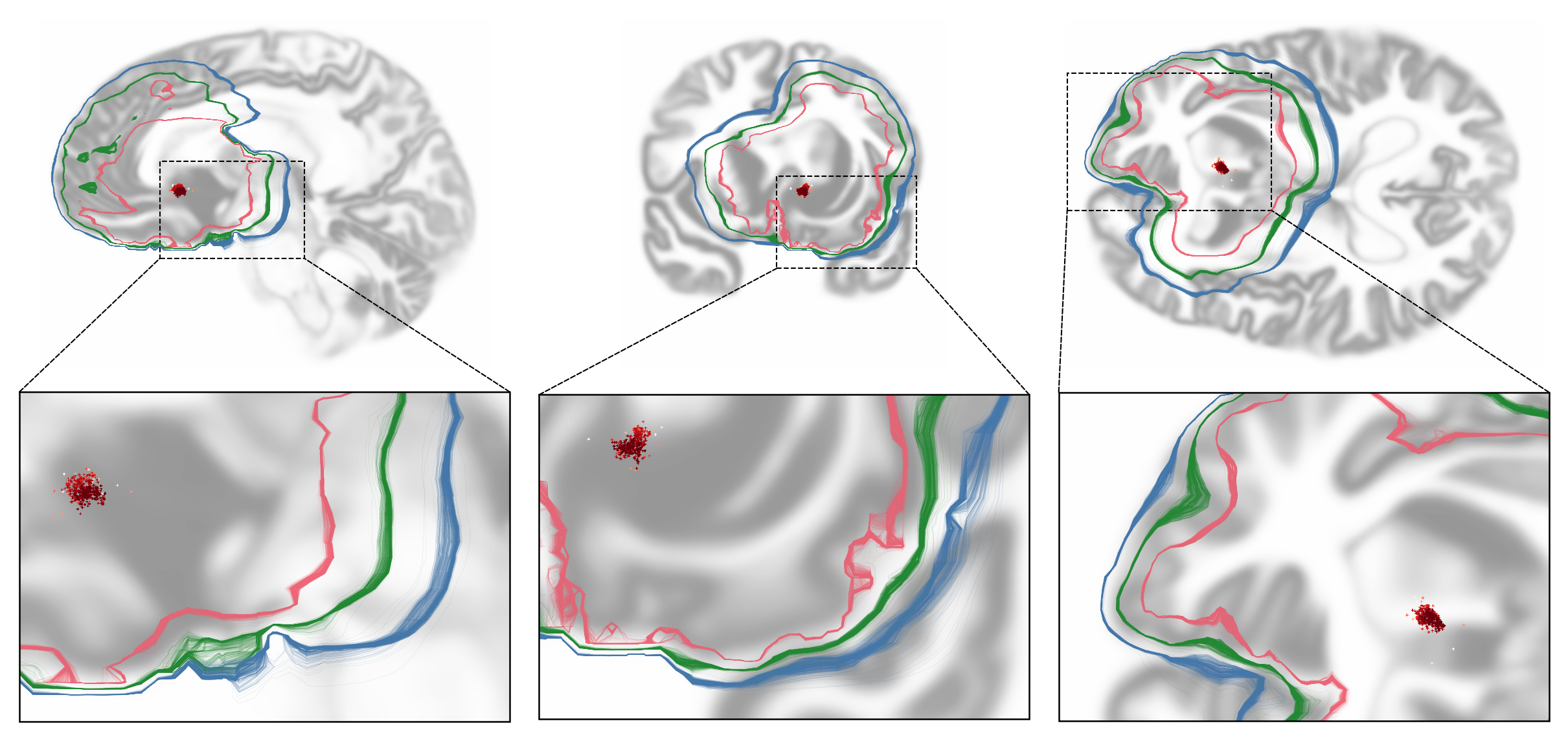}
  \caption{Slice view of the CTVs computed from 256 samples predicted with Gliodil.
    The CTVs correspond to posterior samples of the initial tumor position, and with the same volume as the standard CTV with margins $\SI{1}{\centi\meter}$ (red), $\SI{1.5}{\centi\meter}$ (green) and $\SI{2}{\centi\meter}$ (blue).
    Red symbols show the initial position of the tumor sampled from the posterior.
    Shades of gray indicate the density of gray matter.
    From left to right: sagittal plane, frontal plane, and transverse plane.}
  \label{fig:gliodil:CTV}
\end{figure}

With more than 33.4 millions unknowns, estimating the posterior distribution of the whole field and the model parameters would be prohibitively computationally expensive.
Thus, we once again use the mode approximation described by \cref{eq:posterior:params:approx:MAP}.
Here, we focus on the uncertainties due to the initial tumor position.
The overall log-prior corresponding to this problem is given by
\begin{equation}
  \log P(\mathbf{\theta} | \mathcal{D}) = \argmin_{\mathbf{u}} \left(\lambda_\text{PDE} n_x n_y n_z n_t L_\text{PDE}(\mathbf{u}, \mathbf{\theta}) + \lambda_\text{IC} n_x n_y n_z L_\text{IC}(\mathbf{u}, \mathbf{\theta}) + \mathcal{L}(\mathcal{D}; \mathbf{u}, \mathbf{\theta})\right),
\end{equation}
with $\lambda_\text{PDE}=\SI{e3}{}$, $\lambda_\text{IC}=200$, of the same order of magnitude than in the original GliODIL implementation~\citep{balcerak2023individualizing}, and $L_\text{IC}(\mathbf{u}, \mathbf{\theta})$ is the mean squared error of the residuals between initial conditions from \cref{eq:gliodil:IC} and the solution $\mathbf{u}$ at $t = 0$.

As in the previous section, we use the \ac{BASIS} algorithm to create 512 samples from the posterior distribution $P(x_0, y_0, z_0 | \mathcal{D})$, with $\beta_\text{BASIS}=0.2$, $\gamma_\text{BASIS} = 1.5$, $l_{max}=1$.
For each sample, the optimization is performed over 5000 epochs with the Adam optimizer and a learning rate of $\SI{e-3}{}$, reduced by half if no progress was made for 50 consecutive iterations, with a minimal value of $\SI{e-4}{}$.
Each sample is mapped on a H100 GPU and takes about 15 minutes to evaluate.
We have used \ac{MRI} scans that are part of the dataset used in \citet{balcerak2023individualizing}.

\Cref{fig:gliodil:posterior} shows histograms of these samples.
The initial position inferred by the proposed method is unimodal and concentrated around a position close to the center of mass of the tumor.
The uncertainty of the initial position has a spread of about $\SI{5}{\milli\meter}$, about 5 times the resolution of the \ac{MRI} scan.
This uncertainty is relatively small compared to the size of the tumor.

We now compute the \ac{CTV} associated with each sample.
The \ac{CTV} is defined as the region enclosed with an isosurface of $u$ and with a given volume.
Here we choose volumes of the standard plan with three different margins of $\SI{1}{\centi\meter}$, $\SI{1.5}{\centi\meter}$ and $\SI{2}{\centi\meter}$.
The \ac{CTV} of the standard plan is defined as the region within a given margin distance from the segmented necrotic core.
\Cref{fig:gliodil:CTV} shows slices of the \acp{CTV} obtained with B-ODIL from the top, front and side views of the brain, going through the center of mass of the necrotic core.
We observe a spread of uncertainty of the order of $\SI{1}{\milli\meter}$ to $\SI{5}{\milli\meter}$, depending on the local properties of brain tissues and on the proximity to boundaries.
These findings highlight the ability of the presented Bayesian framework to incorporate spatial uncertainty into \ac{CTV} estimation in a principled and computationally efficient manner.

\section{Summary}
\label{se:summary}

We have introduced B-ODIL, a Bayesian extension to \ac{ODIL}, to solve \ac{PDE}-based inverse problems with quantified uncertainties.
B-ODIL combines prior knowledge in the form of residuals of a \ac{PDE} with the likelihood function of observed data into a posterior density function of possible solutions.
Maximizing this posterior density is equivalent to solving the original \ac{ODIL} method, making the proposed approach consistent with previous studies.
The posterior distribution was estimated through different methods: \ac{HMC} sampling, Laplace approximation, and a mode approximation that allowed one to estimate the parameters of the model in the context of three dimensional \ac{PDE}.
The Laplace and mode approximations gave consistent results with the \ac{HMC} sampling technique, validating their use in higher-dimensional problems where \ac{HMC} is not feasible.

We have applied B-ODIL to four examples of spatiotemporal \acp{PDE} up to four dimensions.
In particular, we have estimated uncertainties for three examples with synthetic data and showed that B-ODIL gave consistent results with the ground truth, with high uncertainties in ambiguous regions such as the ill-posed problem of inferring initial conditions in the diffusion equation.

Finally, we have used B-ODIL to estimate the concentration field of tumor cells in the brain of real patients, by combining data from \ac{MRI} scans with a \ac{PDE} modeling tumor growth.
This application highlights the potential of B-ODIL in a clinical setting, where the quantification of uncertainties is crucial for robust decision-making:
providing a distribution of possible tumor fields, rather than a single-point estimate, can inform clinicians about the confidence in the predictions and guide the design of more effective and personalized treatment plans.
Furthermore, these results demonstrate that B-ODIL can be applied to real-world problems on a large scale and provides a practical framework for solving \ac{PDE}-based inverse problems with quantified uncertainties in science and engineering.

\subsection*{Acknowledgments}

The computations in this paper were run on the FASRC Cannon cluster supported by the FAS Division of Science Research Computing Group at Harvard University.
We thank Sebastian Kaltenbach (Harvard University) and Michal Balcerak (University of Z\"urich) for their valuable input about Bayesian inference and tumor growth model.
This work was supported by the Harvard Data Science Initiative (HDSI) and Amazon Web Services (AWS) under Award Number A59479.

\appendix

\section{Discretization used in GliODIL}
\label{app:gliodil}

\Cref{eq:gliodil:PDE} is discretized on a uniform grid with field values $u_{i,j,k}^n$, where $n$ are time indices and $i$, $j$, and $k$ are spatial indices.
The diffusion and reaction terms are defined as
\begin{equation}
  \begin{aligned}
    A_{i,j,k}^n &= \frac{1}{\Delta x^2} \left(D_{i+\frac 1 2, j, k}^n \left(u_{i+1,j,k}^n - u_{i,j,k}^n\right) - D_{i-\frac 1 2, j, k}^n \left(u_{i,j,k}^n - u_{i-1,j,k}^n\right)\right) \\
    &+ \frac{1}{\Delta y^2} \left(D_{i, j+\frac 1 2, k}^n \left(u_{i,j+1,k}^n - u_{i,j,k}^n\right) - D_{i, j-\frac 1 2, k}^n \left(u_{i,j,k}^n - u_{i,j-1,k}^n\right)\right) \\
    &+ \frac{1}{\Delta z^2} \left(D_{i, j, k+\frac 1 2}^n \left(u_{i,j,k+1}^n - u_{i,j,k}^n\right) - D_{i, j, k-\frac 1 2}^n \left(u_{i,j,k}^n - u_{i,j,k-1}^n\right)\right),
  \end{aligned}
\end{equation}
and
\begin{equation}
  B_{i,j,k}^n = \rho u_{i,j,k}^n \left(1 - u_{i,j,k}^n\right),
\end{equation}
where half indices correspond to the average value of the two adjacent nodes, e.g. \[D_{i+\frac 1 2, j, k}^n = \frac 1 2 \left(D_{i, j, k}^n + D_{i+1, j, k}^n\right).\]
The residuals of the discretized equation are computed from the Crank-Nicolson scheme:
\begin{equation}
  r_{i,j,k}^n = \frac{u_{i,j,k}^{n+1} - u_{i,j,k}^n}{\Delta t} - \frac{A_{i,j,k}^n + A_{i,j,k}^{n+1}}{2} - \frac{B_{i,j,k}^n + B_{i,j,k}^{n+1}}{2}.
\end{equation}

\section{Log likelihood used in GliODIL}
\label{app:gliodil:loglike}

In \cref{se:gliodil}, we have used the log-likelihood described by \cref{eq:gliodil:loglike}, consistent with the original contribution of GliODIL~\citep{balcerak2023individualizing}.
Nevertheless, this formulation may appear somewhat arbitrary from a Bayesian inference perspective.
Here we show that this formulation is in fact closely related to the likelihood used in \cref{se:reaction:diffusion} for small enough values of $\sigma$.

\begin{figure}
  \centering
  \includegraphics{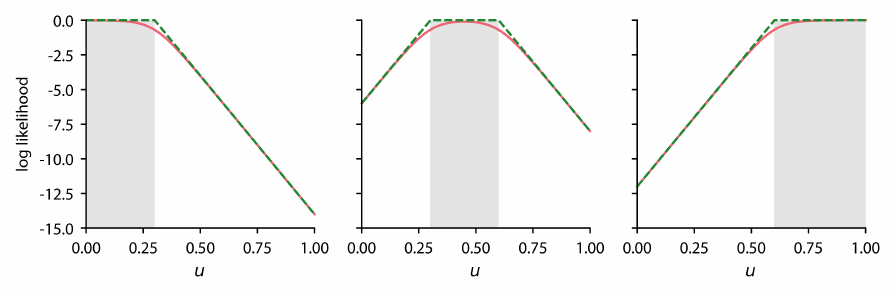}
  \caption{
    Log-likelihood for a single voxel to be classified as healthy (left), glioma (middle), or necrotic core (right), against $u$, with $(\tau_\text{lo}, \tau_\text{hi}) = (0.3, 0.6)$ and $\sigma=0.05$.
    Shaded regions show the intervals $(\tau_{ijk,\text{lo}}, \tau_{ijk,\text{hi}})$.
    Dashed line is the log-likelihood computed in GliODIL (see \cref{eq:gliodil:loglike}), and the solid line represents the log-likelihood described by \cref{eq:gliodil:loglike:nonapprox}.
  }
  \label{fig:gliodil:loglike:comp}
\end{figure}

We generalize the likelihood described in \cref{se:reaction:diffusion} to the case of three classes in the segmented data: healthy, glioma, and necrotic core.
Each class corresponds to a range of tumor concentration.
We consider that voxels are independent from each other, and that the probability of each class is given by
\begin{equation} \label{eq:gliodil:loglike:nonapprox}
  \begin{aligned}
    P(\text{healthy} | u) &= \frac{1}{Z(u)} S\left(\frac{\tau_\text{lo} - u}{\sigma}\right), \\
    P(\text{glioma} | u)  &= \frac{1}{Z(u)} S\left(\frac{\tau_\text{up} - u}{\sigma}\right) \cdot S\left(\frac{u - \tau_\text{lo}}{\sigma}\right), \\
    P(\text{necrotic core} | u) &= \frac{1}{Z(u)} S\left(\frac{u - \tau_\text{up}}{\sigma}\right),
  \end{aligned}
\end{equation}
where $S$ is the sigmoid function defined in \cref{eq:reaction:diffusion:model:alpha} and $Z(u)$ is the normalization factor
\[
Z(u) = 1 + S\left(\frac{u - \tau_\text{up}}{\sigma}\right) \cdot S\left(\frac{\tau_\text{lo} - u}{\sigma}\right).
\]
\Cref{fig:gliodil:loglike:comp} compares this expression with \cref{eq:gliodil:loglike} for $\sigma=0.05$.
Both expressions have very similar values, except for a smoother transition near $\tau\text{lo}$ and $\tau_\text{up}$ for \cref{eq:gliodil:loglike:nonapprox}.
In this work we have used the simpler form described by \cref{eq:gliodil:loglike} to be consistent with the original GliODIL formulation.

\bibliographystyle{elsarticle-num-names}
\bibliography{refs}

@article{karnakov2024solving,
  title={Solving inverse problems in physics by optimizing a discrete loss: Fast and accurate learning without neural networks},
  author={Karnakov, Petr and Litvinov, Sergey and Koumoutsakos, Petros},
  journal={PNAS nexus},
  volume={3},
  number={1},
  pages={pgae005},
  year={2024},
  publisher={Oxford University Press US}
}

@article{menze2014multimodal,
  title={The multimodal brain tumor image segmentation benchmark (BRATS)},
  author={Menze, Bjoern H and Jakab, Andras and Bauer, Stefan and Kalpathy-Cramer, Jayashree and Farahani, Keyvan and Kirby, Justin and Burren, Yuliya and Porz, Nicole and Slotboom, Johannes and Wiest, Roland and others},
  journal={IEEE transactions on medical imaging},
  volume={34},
  number={10},
  pages={1993--2024},
  year={2014},
  publisher={IEEE}
}

@article{karnakov2025optimal,
  title={Optimal Navigation in Microfluidics via the Optimization of a Discrete Loss},
  author={Karnakov, Petr and Amoudruz, Lucas and Koumoutsakos, Petros},
  journal={Physical Review Letters},
  volume={134},
  number={4},
  pages={044001},
  year={2025},
  publisher={APS}
}

@article{amoudruz2025contactless,
  title={Contactless Precision Steering of Particles in a Fluid inside a Cube with Rotating Walls},
  author={Amoudruz, Lucas and Karnakov, Petr and Koumoutsakos, Petros},
  journal={Journal of Fluid Mechanics},
  volume={1014},
  number={},
  pages={A15},
  year={2025},
  publisher={Cambridge University Press}
}

@article{yang2021b,
  title={B-PINNs: Bayesian physics-informed neural networks for forward and inverse PDE problems with noisy data},
  author={Yang, Liu and Meng, Xuhui and Karniadakis, George Em},
  journal={Journal of Computational Physics},
  volume={425},
  pages={109913},
  year={2021},
  publisher={Elsevier}
}

@article{economides2021hierarchical,
  title={Hierarchical Bayesian uncertainty quantification for a model of the red blood cell},
  author={Economides, Athena and Arampatzis, Georgios and Alexeev, Dmitry and Litvinov, Sergey and Amoudruz, Lucas and Kulakova, Lina and Papadimitriou, Costas and Koumoutsakos, Petros},
  journal={Physical Review Applied},
  volume={15},
  number={3},
  pages={034062},
  year={2021},
  publisher={APS}
}

@article{amoudruz2023stress,
  title={The stress-free state of human erythrocytes: Data-driven inference of a transferable RBC model},
  author={Amoudruz, Lucas and Economides, Athena and Arampatzis, Georgios and Koumoutsakos, Petros},
  journal={Biophysical Journal},
  volume={122},
  number={8},
  pages={1517--1525},
  year={2023},
  publisher={Elsevier}
}

@article{wu2018bayesian,
  title={Bayesian annealed sequential importance sampling: an unbiased version of transitional Markov chain Monte Carlo},
  author={Wu, Stephen and Angelikopoulos, Panagiotis and Papadimitriou, Costas and Koumoutsakos, Petros},
  journal={ASCE-ASME Journal of Risk and Uncertainty in Engineering Systems, Part B: Mechanical Engineering},
  volume={4},
  number={1},
  pages={011008},
  year={2018},
  publisher={American Society of Mechanical Engineers}
}

@article{nair2025pinns,
  title={E-PINNs: Epistemic Physics-Informed Neural Networks},
  author={Nair, Ashish S and Jacob, Bruno and Howard, Amanda A and Drgona, Jan and Stinis, Panos},
  journal={arXiv preprint arXiv:2503.19333},
  year={2025}
}

@article{Galbally2010Nonlinear,
  author  = {Galbally, D. and Fidkowski, K. and Willcox, K. and Ghattas, O.},
  title   = {Nonlinear model reduction for uncertainty quantification in large-scale inverse problems},
  journal = {International Journal for Numerical Methods in Engineering},
  volume  = {81},
  number  = {12},
  pages   = {1581--1608},
  year    = {2010},
  doi     = {10.1002/nme.2778},
  issn    = {1097-0029},
}

@article{Lieberman2010Parameter,
  author  = {Lieberman, Chad and Willcox, Karen and Ghattas, Omar},
  title   = {Parameter and State Model Reduction for Large-Scale Statistical Inverse Problems},
  journal = {SIAM Journal on Scientific Computing},
  volume  = {32},
  number  = {5},
  pages   = {2523--2542},
  year    = {2010},
  doi     = {10.1137/090775622},
  issn    = {1064-8275},
}

@article{Ghattas2021Learning,
  author  = {Ghattas, Omar and Willcox, Karen},
  title   = {Learning physics-based models from data: perspectives from inverse problems and model reduction},
  journal = {Acta Numerica},
  volume  = {30},
  pages   = {446--551},
  year    = {2021},
  doi     = {10.1017/S0962492921000064},
  publisher = {Cambridge University Press},
}

@article{raissi2019physics,
  title={Physics-informed neural networks: A deep learning framework for solving forward and inverse problems involving nonlinear partial differential equations},
  author={Raissi, Maziar and Perdikaris, Paris and Karniadakis, George E},
  journal={Journal of Computational physics},
  volume={378},
  pages={686--707},
  year={2019},
  publisher={Elsevier}
}

@article{balcerak2023individualizing,
  title={Individualizing glioma radiotherapy planning by optimization of data and physics-informed discrete loss},
  author={Balcerak, Michal and Weidner, Jonas and Karnakov, Petr and Ezhov, Ivan and Litvinov, Sergey and Koumoutsakos, Petros and Zhang, Ray Zirui and Lowengrub, John S and Wiestler, Bene and Menze, Bjoern},
  journal={Nature Communications},
  year={2025},
  volume={16},
  pages={5982},
  publisher={Nature Publishing Group UK London}
}

@article{balcerak2024physics,
  title={Physics-regularized multi-modal image assimilation for brain tumor localization},
  author={Balcerak, Michal and Amiranashvili, Tamaz and Wagner, Andreas and Weidner, Jonas and Karnakov, Petr and Paetzold, Johannes C and Ezhov, Ivan and Koumoutsakos, Petros and Wiestler, Benedikt and others},
  journal={Advances in Neural Information Processing Systems},
  volume={37},
  pages={41909--41933},
  year={2024}
}

@article{buhendwa2025data,
  title={Data-driven shape inference in three-dimensional steady-state supersonic flows: Optimizing a discrete loss with JAX-Fluids},
  author={Buhendwa, Aaron B and Bezgin, Deniz A and Karnakov, Petr and Adams, Nikolaus A and Koumoutsakos, Petros},
  journal={Physical Review Fluids},
  volume={10},
  number={8},
  pages={084902},
  year={2025},
  publisher={APS}
}

@article{cai2021physics,
  title={Physics-informed neural networks (PINNs) for fluid mechanics: A review},
  author={Cai, Shengze and Mao, Zhiping and Wang, Zhicheng and Yin, Minglang and Karniadakis, George Em},
  journal={Acta Mechanica Sinica},
  volume={37},
  number={12},
  pages={1727--1738},
  year={2021},
  publisher={Springer}
}

@article{karnakov2023flow,
  title={Flow reconstruction by multiresolution optimization of a discrete loss with automatic differentiation},
  author={Karnakov, Petr and Litvinov, Sergey and Koumoutsakos, Petros},
  journal={The European Physical Journal E},
  volume={46},
  number={7},
  pages={59},
  year={2023},
  publisher={Springer}
}

@article{gunes2008use,
  title={On the use of kriging for enhanced data reconstruction in a separated transitional flat-plate boundary layer},
  author={Gunes, Hasan and Rist, Ulrich},
  journal={Physics of Fluids},
  volume={20},
  number={10},
  year={2008},
  publisher={AIP Publishing}
}

@book{suetens2017fundamentals,
  title={Fundamentals of medical imaging},
  author={Suetens, Paul},
  year={2017},
  publisher={Cambridge university press}
}

@article{carrassi2018data,
  title={Data assimilation in the geosciences: An overview of methods, issues, and perspectives},
  author={Carrassi, Alberto and Bocquet, Marc and Bertino, Laurent and Evensen, Geir},
  journal={Wiley Interdisciplinary Reviews: Climate Change},
  volume={9},
  number={5},
  pages={e535},
  year={2018},
  publisher={Wiley Online Library}
}

@article{van2015penalty,
  title={A penalty method for PDE-constrained optimization in inverse problems},
  author={van Leeuwen, Tristan and Herrmann, Felix J},
  journal={Inverse Problems},
  volume={32},
  number={1},
  pages={015007},
  year={2015},
  publisher={IOP Publishing}
}

@article{stuart2010inverse,
  title={Inverse problems: a Bayesian perspective},
  author={Stuart, Andrew M},
  journal={Acta numerica},
  volume={19},
  pages={451--559},
  year={2010},
  publisher={Cambridge University Press}
}

@article{cotter2013mcmc,
  title={MCMC methods for functions: modifying old algorithms to make them faster},
  author={Cotter, Simon L and Roberts, Gareth O and Stuart, Andrew M and White, David},
  journal={Statistical Science},
  pages={424--446},
  year={2013},
  publisher={JSTOR}
}

@book{isakov2006inverse,
  title={Inverse problems for partial differential equations},
  author={Isakov, Victor},
  year={2006},
  publisher={Springer}
}

@article{iglesias2016regularizing,
  title={A regularizing iterative ensemble Kalman method for PDE-constrained inverse problems},
  author={Iglesias, Marco A},
  journal={Inverse Problems},
  volume={32},
  number={2},
  pages={025002},
  year={2016},
  publisher={IOP Publishing}
}

@book{tarantola2005inverse,
  title={Inverse problem theory and methods for model parameter estimation},
  author={Tarantola, Albert},
  year={2005},
  publisher={SIAM}
}

@inproceedings{scott2024using,
  title={Using Scientific Machine Learning to Enhance Measurements of 2{D} Turbulent Flows with {mODIL}},
  author={Scott, Christopher},
  booktitle={2024 Regional Student Conferences},
  pages={90557},
  year={2024}
}

@article{kaltenbach2022semi,
  title={Semi-supervised invertible deeponets for bayesian inverse problems},
  author={Kaltenbach, Sebastian and Perdikaris, Paris and Koutsourelakis, Phaedon-Stelios},
  journal={stat},
  volume={1050},
  pages={8},
  year={2022}
}

@article{lu2021learning,
  title={Learning nonlinear operators via DeepONet based on the universal approximation theorem of operators},
  author={Lu, Lu and Jin, Pengzhan and Pang, Guofei and Zhang, Zhongqiang and Karniadakis, George Em},
  journal={Nature machine intelligence},
  volume={3},
  number={3},
  pages={218--229},
  year={2021},
  publisher={Nature Publishing Group UK London}
}

@article{li2020fourier,
  title={Fourier neural operator for parametric partial differential equations},
  author={Li, Zongyi and Kovachki, Nikola and Azizzadenesheli, Kamyar and Liu, Burigede and Bhattacharya, Kaushik and Stuart, Andrew and Anandkumar, Anima},
  journal={arXiv preprint arXiv:2010.08895},
  year={2020}
}

@article{zou2024neuraluq,
  title={NeuralUQ: A comprehensive library for uncertainty quantification in neural differential equations and operators},
  author={Zou, Zongren and Meng, Xuhui and Psaros, Apostolos F and Karniadakis, George E},
  journal={SIAM Review},
  volume={66},
  number={1},
  pages={161--190},
  year={2024},
  publisher={SIAM}
}

@article{molinaro2023neural,
  title={Neural inverse operators for solving PDE inverse problems},
  author={Molinaro, Roberto and Yang, Yunan and Engquist, Bj{\"o}rn and Mishra, Siddhartha},
  journal={arXiv preprint arXiv:2301.11167},
  year={2023}
}

@book{scherzer2009variational,
  title={Variational methods in imaging},
  author={Scherzer, Otmar and Grasmair, Markus and Grossauer, Harald and Haltmeier, Markus and Lenzen, Frank},
  volume={167},
  year={2009},
  publisher={Springer}
}

@article{gao2022physics,
  title={Physics-informed graph neural Galerkin networks: A unified framework for solving PDE-governed forward and inverse problems},
  author={Gao, Han and Zahr, Matthew J and Wang, Jian-Xun},
  journal={Computer Methods in Applied Mechanics and Engineering},
  volume={390},
  pages={114502},
  year={2022},
  publisher={Elsevier}
}

@article{duthe2025graph,
  title={Graph Transformers for inverse physics: reconstructing flows around arbitrary 2D airfoils},
  author={Duth{\'e}, Gregory and Abdallah, Imad and Chatzi, Eleni},
  journal={arXiv preprint arXiv:2501.17081},
  year={2025}
}

@article{adler2018deep,
  title={Deep bayesian inversion},
  author={Adler, Jonas and {\"O}ktem, Ozan},
  journal={arXiv preprint arXiv:1811.05910},
  year={2018}
}

@article{ding2023full,
  title={Full-volume 3D fluid flow reconstruction with light field PIV},
  author={Ding, Yuqi and Li, Zhong and Chen, Zhang and Ji, Yu and Yu, Jingyi and Ye, Jinwei},
  journal={IEEE transactions on pattern analysis and machine intelligence},
  volume={45},
  number={7},
  pages={8405--8418},
  year={2023},
  publisher={IEEE}
}

@article{angelikopoulos2012bayesian,
  title={Bayesian uncertainty quantification and propagation in molecular dynamics simulations: a high performance computing framework},
  author={Angelikopoulos, Panagiotis and Papadimitriou, Costas and Koumoutsakos, Petros},
  journal={The Journal of chemical physics},
  volume={137},
  number={14},
  year={2012},
  publisher={AIP Publishing}
}

@article{jin2017deep,
  title={Deep convolutional neural network for inverse problems in imaging},
  author={Jin, Kyong Hwan and McCann, Michael T and Froustey, Emmanuel and Unser, Michael},
  journal={IEEE transactions on image processing},
  volume={26},
  number={9},
  pages={4509--4522},
  year={2017},
  publisher={IEEE}
}

@article{kingma2014adam,
  title={Adam: A method for stochastic optimization},
  author={Kingma, Diederik P},
  journal={arXiv preprint arXiv:1412.6980},
  year={2014}
}

@article{liu1989limited,
  title={On the limited memory BFGS method for large scale optimization},
  author={Liu, Dong C and Nocedal, Jorge},
  journal={Mathematical programming},
  volume={45},
  number={1},
  pages={503--528},
  year={1989},
  publisher={Springer}
}

@article{neal2011mcmc,
  title={MCMC using Hamiltonian dynamics},
  author={Neal, Radford M and others},
  journal={Handbook of markov chain monte carlo},
  volume={2},
  number={11},
  pages={2},
  year={2011},
  publisher={Chapman and Hall/CRC}
}

@article{paszke2019pytorch,
  title={Pytorch: An imperative style, high-performance deep learning library},
  author={Paszke, Adam and Gross, Sam and Massa, Francisco and Lerer, Adam and Bradbury, James and Chanan, Gregory and Killeen, Trevor and Lin, Zeming and Gimelshein, Natalia and Antiga, Luca and others},
  journal={Advances in neural information processing systems},
  volume={32},
  year={2019}
}

@article{tierney1986accurate,
  title={Accurate approximations for posterior moments and marginal densities},
  author={Tierney, Luke and Kadane, Joseph B},
  journal={Journal of the american statistical association},
  volume={81},
  number={393},
  pages={82--86},
  year={1986},
  publisher={Taylor \& Francis}
}

@article{clatz2005realistic,
  title={Realistic simulation of the 3-D growth of brain tumors in MR images coupling diffusion with biomechanical deformation},
  author={Clatz, Olivier and Sermesant, Maxime and Bondiau, P-Y and Delingette, Herv{\'e} and Warfield, Simon K and Malandain, Gr{\'e}goire and Ayache, Nicholas},
  journal={IEEE transactions on medical imaging},
  volume={24},
  number={10},
  pages={1334--1346},
  year={2005},
  publisher={IEEE}
}

@article{hogea2007robust,
  title={A robust framework for soft tissue simulations with application to modeling brain tumor mass effect in 3D MR images},
  author={Hogea, Cosmina and Biros, George and Abraham, Feby and Davatzikos, Christos},
  journal={Physics in Medicine \& Biology},
  volume={52},
  number={23},
  pages={6893},
  year={2007},
  publisher={IOP Publishing}
}

@article{hogea2008image,
  title={An image-driven parameter estimation problem for a reaction--diffusion glioma growth model with mass effects},
  author={Hogea, Cosmina and Davatzikos, Christos and Biros, George},
  journal={Journal of mathematical biology},
  volume={56},
  number={6},
  pages={793--825},
  year={2008},
  publisher={Springer}
}

@article{swanson2002quantifying,
  title={Quantifying efficacy of chemotherapy of brain tumors with homogeneous and heterogeneous drug delivery},
  author={Swanson, Kristin R and Alvord Jr, Ellsworth C and Murray, JD},
  journal={Acta biotheoretica},
  volume={50},
  number={4},
  pages={223--237},
  year={2002},
  publisher={Springer}
}

@book{biegler2010large,
  title={Large-scale inverse problems and quantification of uncertainty},
  author={Biegler, Lorenz and Biros, George and Ghattas, Omar and Heinkenschloss, Matthias and Keyes, David and Mallick, Bani and Tenorio, Luis and van Bloemen Waanders, Bart and Willcox, Karen and Marzouk, Youssef},
  volume={712},
  year={2010},
  publisher={Wiley Online Library}
}

@article{miniere2025data,
  title={A data assimilation framework for predicting the spatiotemporal response of high-grade gliomas to chemoradiation},
  author={Miniere, Hugo JM and Hormuth, David A and Lima, Ernesto ABF and Farhat, Maguy and Panthi, Bikash and Langshaw, Holly and Shanker, Mihir D and Talpur, Wasif and Thrower, Sara and Goldman, Jodi and others},
  journal={BMC cancer},
  volume={25},
  number={1},
  pages={1239},
  year={2025},
  publisher={Springer}
}

@article{lamonica2025investigating,
  title={Investigating the Limits of Predictability of Magnetic Resonance Imaging-Based Mathematical Models of Tumor Growth},
  author={LaMonica, Megan F and Yankeelov, Thomas E and Hormuth, David A},
  journal={Cancers},
  volume={17},
  number={20},
  pages={3361},
  year={2025},
  publisher={MDPI}
}

@article{lipkova2019personalized,
  title={Personalized radiotherapy design for glioblastoma: integrating mathematical tumor models, multimodal scans, and Bayesian inference},
  author={Lipkov{\'a}, Jana and Angelikopoulos, Panagiotis and Wu, Stephen and Alberts, Esther and Wiestler, Benedikt and Diehl, Christian and Preibisch, Christine and Pyka, Thomas and Combs, Stephanie E and Hadjidoukas, Panagiotis and others},
  journal={IEEE transactions on medical imaging},
  volume={38},
  number={8},
  pages={1875--1884},
  year={2019},
  publisher={IEEE}
}

@article{bui2013computational,
  title={A computational framework for infinite-dimensional Bayesian inverse problems Part I: The linearized case, with application to global seismic inversion},
  author={Bui-Thanh, Tan and Ghattas, Omar and Martin, James and Stadler, Georg},
  journal={SIAM Journal on Scientific Computing},
  volume={35},
  number={6},
  pages={A2494--A2523},
  year={2013},
  publisher={SIAM}
}

@article{yu2025conformal,
  title={A conformal prediction framework for uncertainty quantification in physics-informed neural networks},
  author={Yu, Yifan and Ho, Cheuk Hin and Wang, Yangshuai},
  journal={arXiv preprint arXiv:2509.13717},
  year={2025}
}

@article{petra2014computational,
  title={A computational framework for infinite-dimensional Bayesian inverse problems, Part II: Stochastic Newton MCMC with application to ice sheet flow inverse problems},
  author={Petra, Noemi and Martin, James and Stadler, Georg and Ghattas, Omar},
  journal={SIAM Journal on Scientific Computing},
  volume={36},
  number={4},
  pages={A1525--A1555},
  year={2014},
  publisher={SIAM}
}

@article{antil2024efficient,
  title={Efficient algorithms for Bayesian inverse problems with Whittle--Mat{\'e}rn priors},
  author={Antil, Harbir and Saibaba, Arvind K},
  journal={SIAM Journal on Scientific Computing},
  volume={46},
  number={2},
  pages={S176--S198},
  year={2024},
  publisher={SIAM}
}

@article{wang2023stochastic,
  title={Stochastic modeling and statistical calibration with model error and scarce data},
  author={Wang, Zhiheng and Ghanem, Roger},
  journal={Computer Methods in Applied Mechanics and Engineering},
  volume={416},
  pages={116339},
  year={2023},
  publisher={Elsevier}
}

@article{morrison2018representing,
  title={Representing model inadequacy: A stochastic operator approach},
  author={Morrison, Rebecca E and Oliver, Todd A and Moser, Robert D},
  journal={SIAM/ASA Journal on Uncertainty Quantification},
  volume={6},
  number={2},
  pages={457--496},
  year={2018},
  publisher={SIAM}
}

\end{document}